\documentclass{article}
\usepackage[numbers]{natbib}
\usepackage{lineno,hyperref}
\usepackage{subfigure}
\usepackage{graphicx}
\usepackage{authblk}
\usepackage{setspace}
\usepackage{moreverb,url}
\usepackage{subfigure}
\usepackage{graphicx}
\usepackage{times} 
\usepackage{microtype}
\usepackage{amssymb}
\usepackage[
left=4cm,
right=4cm,
top=4cm,
bottom=4cm
]{geometry}
\title{Closure under Coupling of Cellular-DEVS for the Optimization of Memory Resource: Wildfire Spread Case Study}

\author[1]{Youcef Dahmani}
\author[2]{Maamar El Amine Hamri}
\author[3]{Nesrine Driouche}
\affil[1]{Ibn Khaldoun University, Tiaret, Algeria (dahmani\_youcef@yahoo.fr)}
\affil[2]{Aix-Marseille Univ, Université de Toulon, CNRS
	LIS, Marseille, France (amine.hamri@lis-lab.fr)}
\affil[3]{Independant software researcher, Marseille, France (driouche.nesrine@hotmail.fr)}

\begin{document}

\maketitle

\begin{abstract}
The present work aims to show one of the advantages of using the property of closure under coupling in the DEVS specification. The advantage concerned in this paper attempts to address the need for memory resources during the simulation of systems by cellular-DEVS.

This improvement of performance is based on the usage of the property closure under coupling in the DEVS formalism. With this property and taking account of the iterative behavior of each cellular-DEVS atomic model, we provide simulation of many models simultaneously. The method starts with the specification of the cellular-DEVS coupled model which is then converted into its equivalent DEVS atomic model. Thus, the goal of this conversion is to transform large quantities of atomic models coupled together, which require huge computational resources, into one DEVS atomic model. 

A case study is presented at the end of the work on modeling and simulation of forest fire propagation using DEVS and cellular-DEVS. A specification by cellular-DEVS of the forest fire model and its non-modular equivalent DEVS atomic model are presented. Finally, a comparison between both methods is presented in term of consumption of resources.
\end{abstract}

\section{Introduction}
\label{intro}
%Discrete event Modeling and Simulation (M\&S) is one of the most popular paradigms to model dynamic systems  \cite{Banks2000}. Both scientists and industrials have defined methodologies, approaches and formalisms to assist users (experts in a specific domain) to define and simulate models.

Modeling and Simulation (M\&S) have accumulated a large number of successes in a wide and varied range of domains. However, M\&S consumes time, effort and resources when you need to resolve class of problems where the analytic or modeling solution is very hard to find. This class of problems includes dynamic systems where there is a wide number of parameters with time and causal dependencies between them, a non-linear behavior of the system, etc.

Discrete event M\&S deals with systems whose temporal and spatial behaviors are complex to be treated analytically. The DEVS formalism (Discrete EVent system Specification) is one of the common formalism used in the simulation of dynamic systems \cite{ Zeigler1976, Zeigler2000}.  It is known for its modularity and expressiveness. This formalism offers, compared to others, a general purpose.

However, the DEVS formalism has undergone several extensions to meet specific needs. Many variants on DEVS were adopted by introducing appropriate theories such the parallel-DEVS \cite{Chow1996} DS-DEVS \cite{Barros1997}, Cell-DEVS \cite{Wainer2001}, etc.

Although these variants exist, it is very difficult to simulate complex systems, especially those represented by large-scale cellular-DEVS models. The need for computational resources becomes vital in order to ensure reliable and fast simulations. Many techniques were adopted with their advantages and disadvantages \cite{Shiginah2011}.

In this work, we take advantage of the property, closure under coupling of DEVS which allows us to reduce computational resources without changing the bases of DEVS specification.

A case study of the Wildfire Spread Simulation is modeled by the cellular-DEVS coupled model and its equivalent DEVS atomic model. Simulation results for both techniques are given and a comparison between these two simulations is described.

The remainder of this paper is outlined as follows: Section~\ref{stateDEVS} resumes the basic concepts on the DEVS formalism. Section~\ref{sec:relatedworks} is divided into two subsections, in the first one, a literature review on optimization methods to improve the performances of simulation within the DEVS formalism is reported, and in the second subsection, an overview on forest fire spread simulations is presented. Section ~\ref{sec:formalspec} depicts our approach by converting the cellular-DEVS into its equivalent DEVS atomic model using the property closure under coupling applied to wildfire spread. Section~\ref{sec:experiments} illustrates simulation experiments on our case study on forest fire spread, we specify both models: cellular-DEVS coupled model versus DEVS atomic model.  At the end, conclusion and perspectives are given.

\section{DEVS Formalism}
\label{stateDEVS}
\subsection{Background}
\label{DEVSform}

Being based on system theory, the DEVS formalism is one of the means theoretically well-grounded to express discrete event systems in a hierarchical and modular manner. A discrete event system is a dynamic system whose behavior is directed by the appearance of discrete events \cite{Lee2003}. 

The use of discrete event-based simulation, rather than time-driven simulation has been proven to reduce computation time in many applications.  DEVS formalism can be used to specify systems whose input, state and output trajectories are piecewise constant ~\cite{Zeigler1995}. In addition, DEVS can give highest performance for simulation of continuous systems typified by spatial and temporal heterogeneity \cite{Hu2004, Nutaro2003}.

Theoretically, each system is characterized by two features: functional (behavioral) and structural aspects. At the lowest level, a basic part called DEVS atomic, describes the autonomous behavior of a discrete event system. At the highest level, DEVS  coupled describes a system as modular and hierarchical structure ~\cite{Zeigler1976, Zeigler2000}.

A DEVS atomic model is based on continuous time, inputs, outputs, states and functions (output, internal and external transitions, life states). Formally, a DEVS atomic model is described by the following seven-tuple:

$ AM_{devs} = (X, S, Y, \delta_{int}, \delta_{ext}, \lambda, t_{a}) $

Where 
$X$, $S$ and $Y$ are the sets of input events, states, and output events, respectively, $ \delta_{int}: S\rightarrow S$ the internal transition function, describes the state changes, that occurs when the elapsed time reaches the lifetime of the state, $\delta_{ext}: Q\times S \rightarrow S$ is the external transition function, where $Q = \{(s,e) | s \in S , e \in {R}^{+}  , 0 \leq e \leq t_{a}(s)\}$ is the  total state set and $e$ describes the elapsed time since the last transition of the current state $s$, $\lambda:S \rightarrow Y$: when the elapsed time reaches the lifetime of the state, this function generates an output event, and $t_{a}: S\rightarrow {R}^+\cup\infty$: time advance function, which is the lifespan of a state.

The DEVS coupled model allows formalizing the modeled system in a set of inter-connected and reused components. A DEVS coupled model is defined as an eight-tuple: 
 
$ CM_{devs} = (X_{self}, Y_{self}, D, \{M_{d}| d \in D\}, EIC, EOC, IC, Select)$

Where $X_{self}$: set of inputs events, $Y_{self}$ : set of outputs events, $D$ : is the name set of sub-components, $\{M_{d}| d \in D\}$:  set of sub-components of DEVS models, $EIC$: set of External Input Coupling, $EOC$: set of External Output Coupling, $IC$: defines the Internal Coupling, and $Select: 2^D \rightarrow D$:  defines a priority between simultaneous events. 

\subsection{Cellular-DEVS }
\label{DEVSCellular}

Several DEVS formalisms have evolved compared to the classic DEVS to fit specific needs. Cellular-DEVS models have been created to model and simulate several phenomena such as fire propagation, traffic control, etc. \cite{Hu2004}.

Cellular-DEVS originates from the cellular automata formalism. The latter is based on discrete-time simulation which consumes the computation power to update all cells at each time-step. A decrease in the time-step used in the models would increase accuracy, but would also result in a longer simulation time and requires large computational resources. In fact, in many cases, there are few cells that are concerned to be updated at each time-step, which makes this manner inefficient. In order to overcome this problem, cellular-DEVS was adopted to offer computational resources to the active cells that really execute state transitions and therefore avoid useless computation on inactive cells. 

The cellular-DEVS formalism divides the spatial space into a set of identical cells where  computations are done locally. A cell is considered as a DEVS atomic model which executes the local computations based on its own state as well as its neighbor states. The spatial space is implemented as a DEVS coupled model where the internal couplings between cells are given by neighborhood rules \cite{Shiginah2011}.

\subsection{Closure under Coupling}
\label{DEVSClosureundercoupling}

The property closure under coupling, in DEVS and parallel-DEVS, reports that every coupled model has its own equivalent atomic model. Therefore a DEVS coupled model regroups several DEVS models, which can be regarded as another DEVS atomic model.

The transition from the parallel-DEVS coupled model into its equivalent parallel-DEVS atomic model is described as follows \cite{Shiginah2006, Shiginah2011}:

The state set $S$ of the subsequent model will be the Cartesian product of the total state sets of all the DEVS atomic models. Thus, the time advance $t_{a}(s)$ defines the time remaining to the next event in component $d$.\\

$S = \times_{d\in D}Q_{d}$\\  							
$t_{a}(s) = minimum\{\rho_{d} | d\in D\}$ \\Where \\
$ s\in S , s = (\dots, (s_{d},e_{d}), \dots) \hbox{ for all } d \in D \hbox{ and } \rho_{d} = t_{a}(s_{d})-e_{d}$\\
The overall transition function is defined as:
\\
\\
 $\delta(s,e,x^{b})= \left\{\begin{array}{ll}
 \delta_{ext}(s,e,x^{b}) &\mbox{ if $0\leq e<t_{a}(s)\hbox{ and } x^{b}\neq\phi $}\\
\delta_{conf}(s,e,x^{b}) & \mbox{ if $e = t_{a}(s) \hbox{ and } x^{b} \neq \phi$}\\
\delta_{int}(s) & \mbox { if $e = t_{a}(s) \hbox{ and } x^{b} = \phi $}
 \end{array}\right.$

Where\\
$\delta_{ext} : \times_{d\in D}Q_{d} \times X \rightarrow \times_{d\in D}Q_{d}$\\
$\delta_{conf} : \times_{d\in D}Q_{d} \times X \rightarrow \times_{d\in D}Q_{d}$\\
$\delta_{int} : \times_{d\in D}Q_{d} \rightarrow \times_{d\in D}Q_{d}$\\
$\lambda : \times_{d\in D}Q_{d} \rightarrow Y$\\

DEVS coupled model can be reduced to a behaviorally equivalent DEVS atomic one. Thus, the concept of closure under coupling ensures that the coupled model results in a model of the same class which has a basic specification.

\section{Related  Works}
\label{sec:relatedworks}
\subsection{Optimization Techniques in DEVS Simulation}
\label{OptTecDEVSim}

In many situations, simulate complex systems with DEVS requires machines with very high performance. To overcome this crucial need, various techniques have been used and can be classified in two categories: those that modify the DEVS formalism by adding either specific functions or variables to manage in a dynamic way the structure of the models, which require demonstrating again some of the properties of DEVS (closure under coupling, hierarchy, etc.), and those that preserve the DEVS formalism but which integrate information concerning the behavioral model structure, which decrease the modularity, flexibility and its reuse.

For both categories, different techniques have been developed in the DEVS community such as quantization \cite{Kofman2001}, which helps in enhancing performance simulation by decreasing the number of state transitions and messages. However, \cite{Shiginah2011} have noted that quantization is an approximation method that satisfies the tolerance requirement, i.e, the simulation should stay within acceptable error.

The parallel and distributed DEVS simulations need more hardware as well as extra work for parallelization of prevailing sequential models \cite{Wang1992,Park2003,Lee2003}. However, \cite{Wainer2010} have noted that, the overhead of check pointing and rollback operations may result in unstable and degraded performance. 

DEVS components within variable structure \cite{Hu2005}, permit  models and  couplings to be dynamically added and/or removed during simulations run. This main advantage is loosen when an important number of update (add and/or remove) of models and couplings are done during a simulation cycle~\cite{Sun2009}. 

However, there are few works based on converting DEVS coupled into DEVS atomic to increase performance \cite{Shiginah2011}. Among these works,  \cite{Lee2003} presented a formal approach of this property with addition of scheduling mechanism, they introduced a composition-based method that converts a DEVS coupled model into its equivalent DEVS atomic model at compile time in order to speed up the simulation by accounting events and messages at compile time. The main advantage is that the resultant atomic model should keep track of all the functions of these internal models. However, this technique results in overhead as the number of the internal models grows up, and the conversion process will be more complex to be done properly. \cite{Beltrame2006}  followed the same approach by converting a coupled model into atomic one in order to eliminate the message overhead based on Modelica parallel variable update. Unfortunately, this approach shows slow simulations because of the large number of variables used. Thereby, these two techniques proceed at compile time to allow the conversion.

In \cite{Shiginah2011}, the idea is to convert a set of cellular-DEVS atomic model into its single DEVS atomic model while keeping the same accuracy and remaining an error-free approach. Thus, instead of considering each cell as an atomic model, with the closure under coupling and at the specification and development stage, the method will transform a group of cells into a non-hierarchal and non-modular model which leads to increase the performance simulations run. 

Although the resulting model is not hierarchical, but this does not present any problem to the user since they are identical sub-models. Therefore, the user will bypass the hierarchy due to the iterative and repetitive behavior and structure of cells.  

The present work is closely related to \cite{Shiginah2011}. Our work focuses mainly on the usage of the memory resource. We apply the property closure under coupling in the modeling of forest fire to obtain reliable simulations by decreasing the use of memory resource; and as a collateral profit, we speed up the simulation run and we increase considerably the number of concurrent simulated cells.

\subsection{ Forest fire spread}
\label{ForFirSpr}

Wildfires continue to cause considerable losses of  human lives, wildlife and houses every year and are costly to contain. In United States of America alone, about one billion dollars is spent annually on wildfire suppression and containment ~\cite{Ntaimo2008,Hu2012}. 

Fire spread is a propagation process and needs to build simulation models that relate the system evolution as accurately as possible to plan scenarios in order to save lives and money. However, simulating wildfire spread remains a challenging problem due to the complexity of wildfire behavior.

Such far, no model has tried to combine the different features of fire behavior  (vegetation, weather, topography, etc.) that are already established separately because of the important factors that influence wildfire behavior and the interaction between them \cite{Finney2004,Ntaimo2008}, that why there are many models in the literature. 

The most common approaches to simulate fire spread are based either on vector or wave approach, or on the cellular models \cite{Finney2004}. For the former, the model produces vector fire perimeters at determined time intervals. For the latter, the cellular models simulate fire spread as a discrete process of ignitions across a regular division of the space in cells. Successive computations are carried out on each cell to ignite its neighbors.

The majority of fire models in use today are principally based on fire propagation relations developed by  \cite{Rothermel1972}, \cite{Albini1976} and  \cite{Anderson1983}. The popular Rothermel model was in particular chosen because of its robustness and stability which have been extensively tested and proven \cite{DAmbrosio2006,Ntaimo2008}. 

Both categories of models (vector, cellular) are simulated either in discrete or continuous time \cite{Ntaimo2006}. The following non-exhaustive list of examples uses the Rothermel model. HFIRE \cite{Morais2001} includes discrete time model, FARSITE \cite{Finney2004} where the fire growth is based on Huygens principle of wave propagation, and BEHAVE \cite{Burgan1984} uses continuous simulation for the vector models. 

The forest fire simulation models include cellular space, cellular automata \cite{DAmbrosio2006}, cellular-DEVS models \cite{Vasconcelos1995}, DEVS-Fire \cite{Ntaimo2004}, and cell-DEVS which uses heat transfer partial differential equations to calculate fire spread in each cell [\cite{Ameghino2001,Muzy2002}].  

Cellular-DEVS and Cell-DEVS simulation results were compared and the analysis concluded the validation of both models \cite{Muzy2002}. However, the present cellular implementation knows a simulation issue whenever all cells in the cell space are created simultaneously which causes simulation performance degradation. This degradation is attributed to the initialization of the cells and the functions needed, and the memory that is required to run the simulation \cite{Ntaimo2008}.

A dynamic structure was proposed to overcome this issue \cite{Ntaimo2008,Hu2012}. This approach keeps track of active cells along the fire front as the simulation proceeds. The cells are created/deleted dynamically; however this implementation has computational overhead at runtime when we proceed with a large-scale cell space.

Other implementations were proposed to overcome these concerns. Indeed, besides using high performance computers, researchers still develop other techniques including innovative algorithms that allow handling only active cells or research on how the non-modularity/non-hierarchy in modeling can improve simulations run \cite{Shiginah2011, Hamri2013}.

\section{Formal Specification and Design Issue}
\label{sec:formalspec}
The M\&S of forest fire spread is confronted to size of the map in a geographic information system (GIS). So, obtaining reliable and accurate simulations needs huge computation performance. The aim of this section is to show the benefit of applying the closure under coupling property of DEVS to enhance the use of memory space and increase the size of the map on GIS during simulation runs.
\subsection{ Problem Description}
\label{PrbDes}

The forest is represented as a 2D cell space of square cells whose dimensions depend on the resolution of the GIS. Each cell has eight neighbors and carried out its local computation of the rate of fire spread and direction based on its local conditions (parameters). The literature classifies the parameters which set the fire spread ratio into three groups: vegetation type (caloric content, density, etc.), fuel properties (vegetation size) and environmental parameters (wind speed, humidity and slope). The flaming fire evolves principally according to the wind speed and its direction \cite{Scot2005}. We assume, in this work,  that fuel, topography and weather conditions are uniform for each cell.

This work is closely related to \cite{Ntaimo2008, Hu2012}. The rate of fire spread of each cell relies on Rothermel model \cite{Rothermel1972} which is decomposed thereafter into eight spreading directions according to the cell neighborhood and schedules ignition of its neighboring cells accordingly. 

The ignition process consists on spreading fire from a burning cell to its neighbors using Moore neighborhood. A cell space is considered as a DEVS atomic model, therefore, the forest cell space is a coupled model composed of a set of cell atomic models.

We consider that each cell can be in one of the following possible states:
\begin{itemize}

\item Nonflammable (N): It can be a road, a surface of water or just an empty surface.
\item Unburned (U): Passive state; it represents any fuel which is not consumed yet by fire.
\item Burning (B): represents a consuming fire.
\item Ash (A): It is afterburning state, it is the final combustion process state.
\end{itemize}

\subsection{Functional Architecture}
\label{sec:funcarchi} 

The overall system proposed in this work is composed of (Figure~\ref{fig:funcarchisim}):
\begin{figure*}[!h]
{
\centering
%\figure[The LSIS\_DME approach.]
{\includegraphics[trim= 0mm 67mm 82mm 0mm, clip, height = 8 cm, width= 12cm]{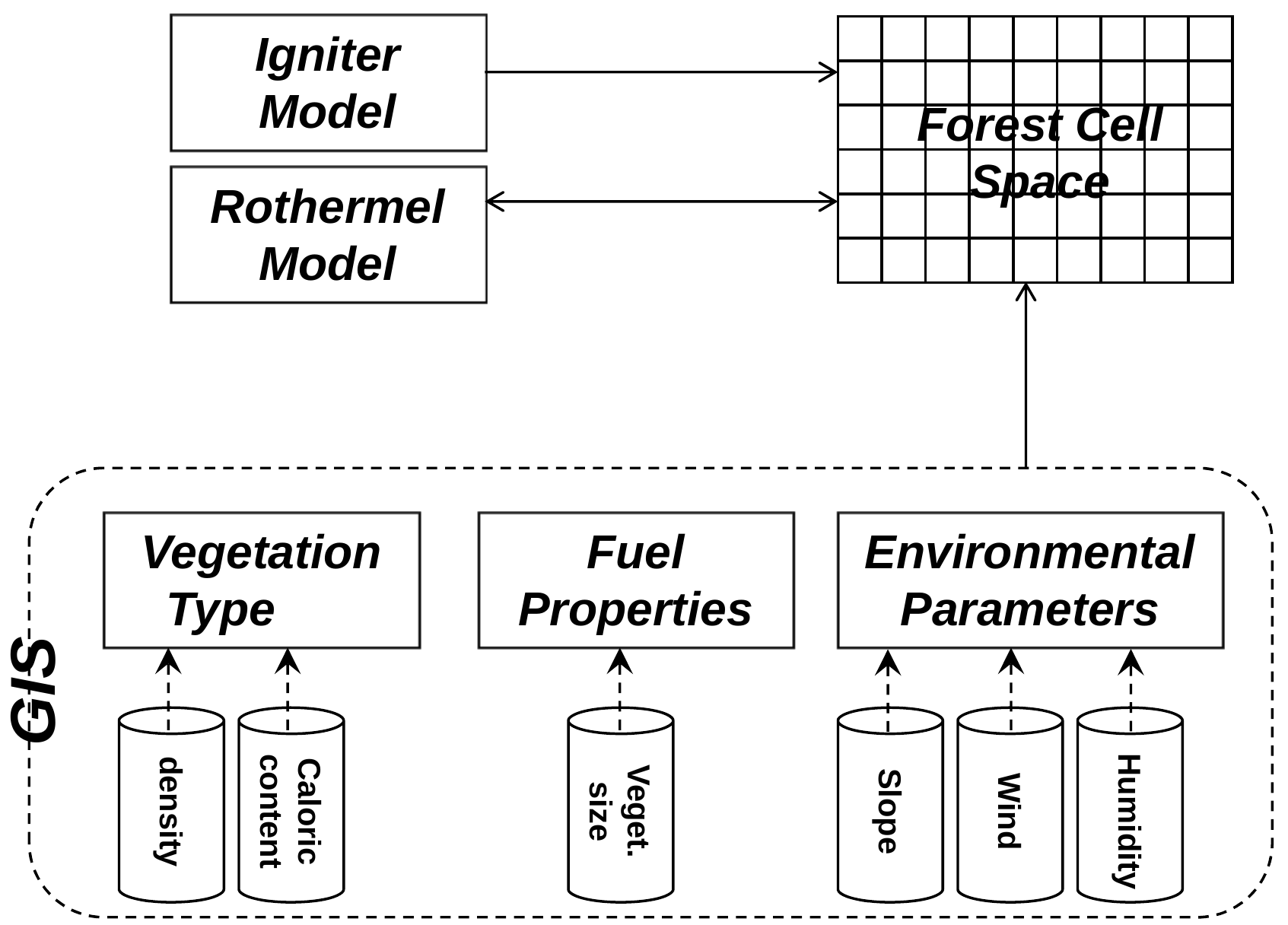}}
%.49\linewidth
\caption{Functional architecture of fire spread simulation.}
\label{fig:funcarchisim}
}
\end{figure*}

\begin{itemize}

\item Forest cell space, in which each forest cell is an atomic model. 
\item Geographic information system interface which provides the different data that influence the spread rate. It consists of: 
\begin{itemize}
\item Environmental parameters interface: provides weather and topography data such as wind speed, its direction, relative humidity, slope, aspect, etc.,
\item Fuel properties: provide size of vegetation, fuel model, etc., and
\item Vegetation type: provides density of vegetation, its calorific content, etc.
\end{itemize}
\item Forest cell igniter: ignites an initial set of cells to start the simulation,~and
\item Rothermel model: computes fire spread and its direction.
\end{itemize}

In its basic form, Forest cell space is modeled as a grid composed of $n$ rows and $m$ columns, which depends on the data resolution. The dynamic system of the flaming front propagation speed is given by the simulator. It is based on the current cell position and its own variables, each of which is given by an appropriate model. The forest cell igniter is a DEVS atomic model; it is coupled to all forest cells to ignite cells. The Environmental interface and Fuel and Vegetation interface provide two kinds of values: spatial-temporal and environmental data. These data are fed into the forest cells which in turn send these data to Rothermel model to compute the rate of fire spread and its main direction.

\subsection{Formal Specification and Implementation}

In this section, we are going to describe the overall system design. As seen above, the fire spread model is composed of forest cell space coupled to the igniter model, Rothermel model, and GIS interface (environmental, vegetation and fuel models).

\subsubsection{Forest cell DEVS atomic model}
\label{sec:forestcellspaceAM}

For each cell $O=(a,b)$, its Moore neighborhood  is given by the set:\\
\hspace*{2cm}$V = \{O, N, NE, E, SE, S, SW, W, NW\} $\\

\begin{figure}[h]
{
\centering
%\figure[The LSIS\_DME approach.]
{\includegraphics[trim= 20mm 88mm 175mm 10mm, clip, height = 4 cm, width= 4cm]{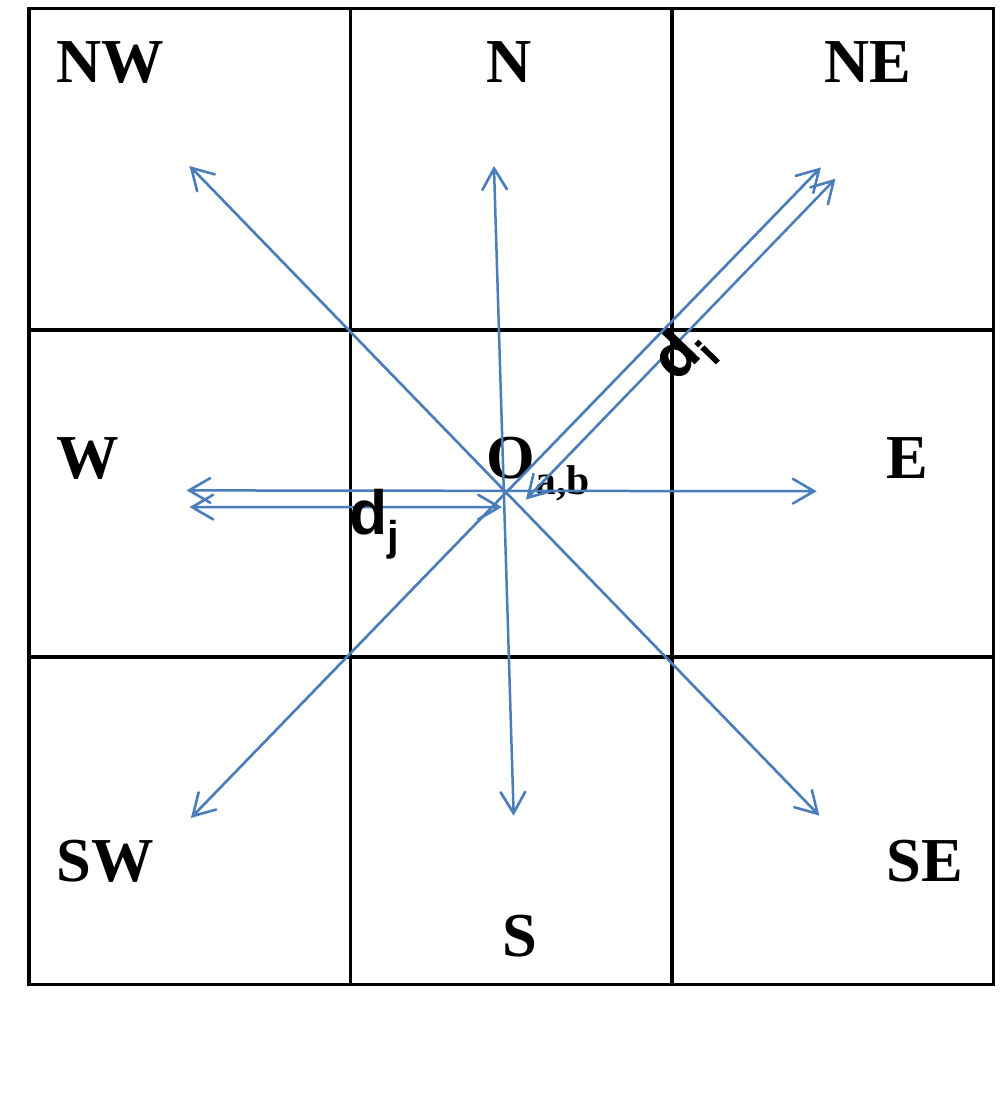}}
%.49\linewidth
\caption{Cell center-to-center fire spread.}
\label{fig:cellcentor-to-center}
}
\end{figure}

Each neighbor is reported as one of eight compass points  $(N, NE, E, SE, $ $S, SW, W, NW)$. In Figure~\ref{fig:cellcentor-to-center}, the fire is propagated from the center of the cell to the center of the neighbor cells.

Each forest cell DEVS atomic model has four states: Nonflammable (N), Unburned (U), Burning (B) and Ash (A). All cells are assumed to be initially in unburned state, except those defined by the user in the passive state Nonflammable.

Once a cell is ignited it computes the ignition delays for its eight neighbors, and sends out the ignition message for each of them. Therefore, each cell is ignited either by its neighbor or by the igniter model. The ignited cell, in its turn, ignites its neighbors. Thus, an additional input port is coupled with each forest cell DEVS atomic model. 
The ignition delays are calculated by the Rothermel model. Each cell sends out its state variables (fuel model, size of vegetation, wind speed, wind direction, etc.) obtained from the GIS interface toward the Rothermel model. The latter sends out as a result the rate of fire spread and its main direction to the concerned cell. Therefore the forest cell interacts with the weather and fuel interface, and the Rothermel model via these additional input/output ports.    

\begin{figure}[h]
{
\centering
%\figure[The LSIS\_DME approach.]
{\includegraphics[trim= 0mm 137mm 177mm 0mm, clip, height = 6 cm, width= 7.5cm]{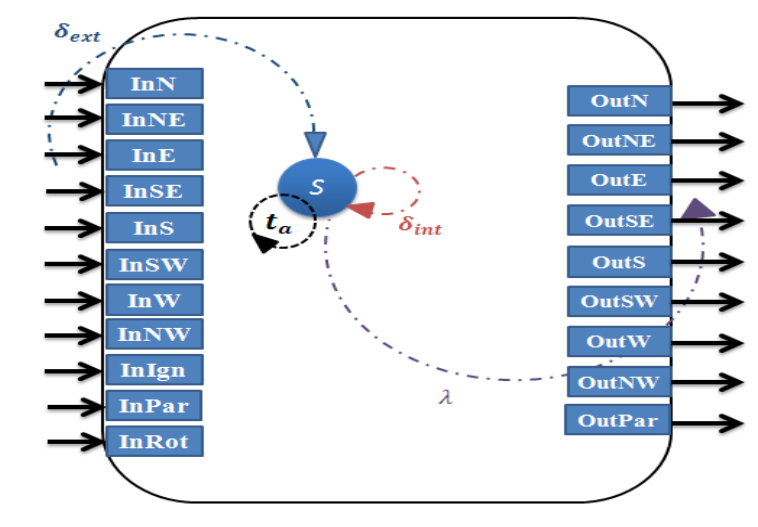}}
%.49\linewidth
\caption{Forest cell DEVS atomic model.}
\label{fig:forestcellDEVSatomic}
}
\end{figure}

Consequently, each forest cell has 11 input ports and 9 output ports by which it reacts and acts on its environment~(Figure~\ref{fig:forestcellDEVSatomic}).

The dynamics of fire spread is described as follows: once a cell is ignited, it sends out ignition messages one by one in each of the eight directions. The decomposition of fire spread in each direction is based on the model that defines the fire shape as an ellipse \cite{Hu2012, Finney2004}. The contagion process of fire spread across a cell considered in our case study is center-to-center as in \cite{Ntaimo2006}. Consequently, the burning cell remains in this state at least until all the eight messages will be sent out. \\
Let us consider: 

\begin{itemize}
\item	$T_{td} = \{t_{1}, t_{2}, t_{3}, t_{4}, t_{5}, t_{6}, t_{7}, t_{8}\}$, the set of these eight different time delays sorted respectively from the smallest to the largest.  
\item $\Delta T_{td} = \{\rho_{1}, \rho_{2}, \rho_{3}, \rho_{4}, \rho_{5}, \rho_{6}, \rho_{7}, \rho_{8}\}$,\\ the set of the time interval between two adjacent ignition messages. \\
$\rho_{1}=t_{1} \hbox{ and }  \rho_{i} = (t_{i}-t_{i-1}) | i=2..8$.
\end{itemize}

\begin{figure}[!h]
{
\centering
%\figure[The LSIS\_DME approach.]
{\includegraphics[trim= 0mm 90mm 140mm 0mm, clip, height = 5 cm, width=6cm]{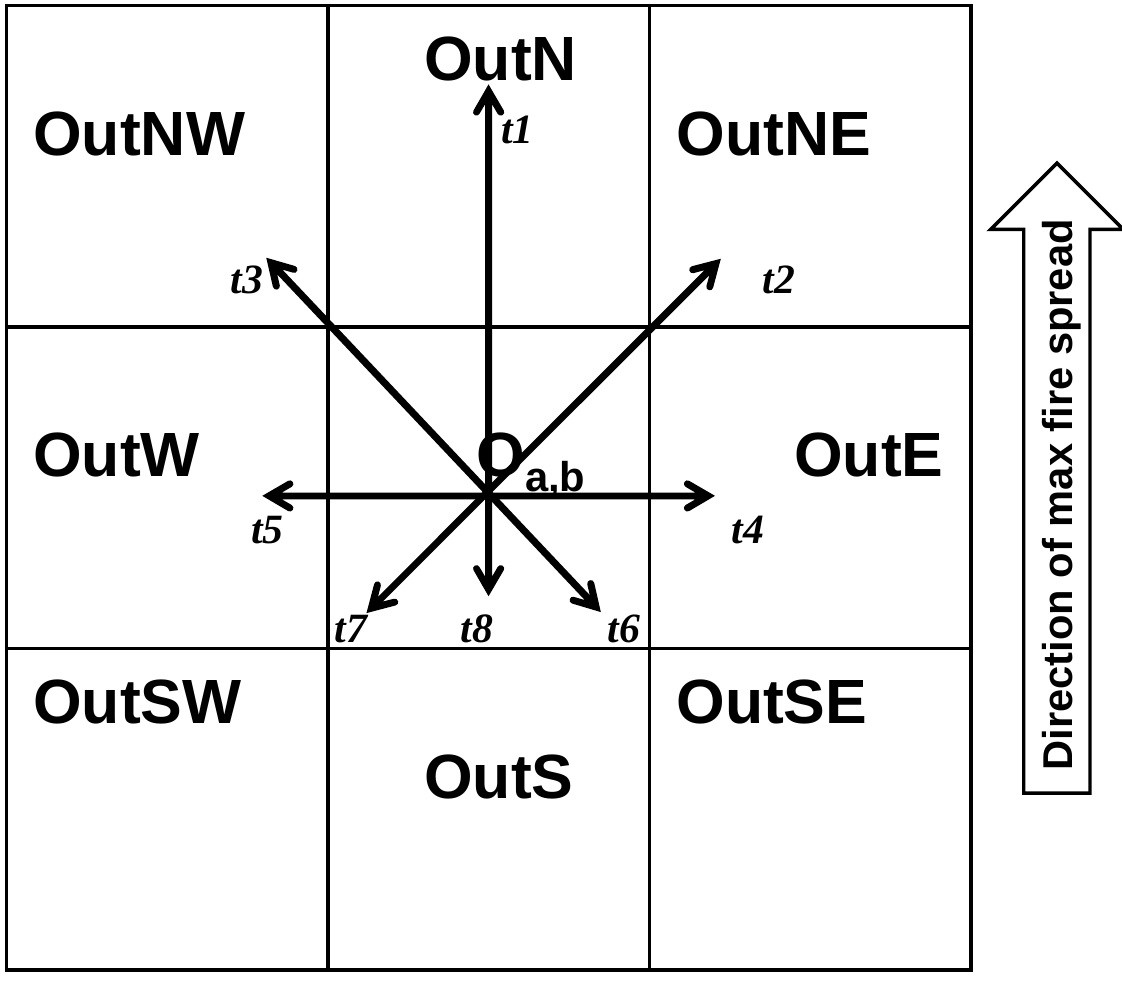}}
%.49\linewidth
\caption{Rate of spread calculation in eight directions.}
\label{fig:ros8Direction}
}
\end{figure}

The first message is sent at $t+t_{1}$ and the last one at $t+t_{8}$, at this moment, the forest cell transitions from the burning state to the ash one (burned) after sending out its last ignition message. The set $T_{td}$ can be affected by the parameters change (wind flow, speed, direction) and consequently the set $\Delta T_{td}$. In this case, the forest cell space model notifies the Rothermel model which updates the rate of fire spread and subsequently the spread in each direction. The rate of fire spread in each direction is depicted on Figure~\ref{fig:ros8Direction}.

\begin{figure}[!h]
{
\centering
%\figure[The LSIS\_DME approach.]
{\includegraphics[trim= 0mm 109mm 84mm 0mm, clip, height = 5 cm, width=7.5cm]{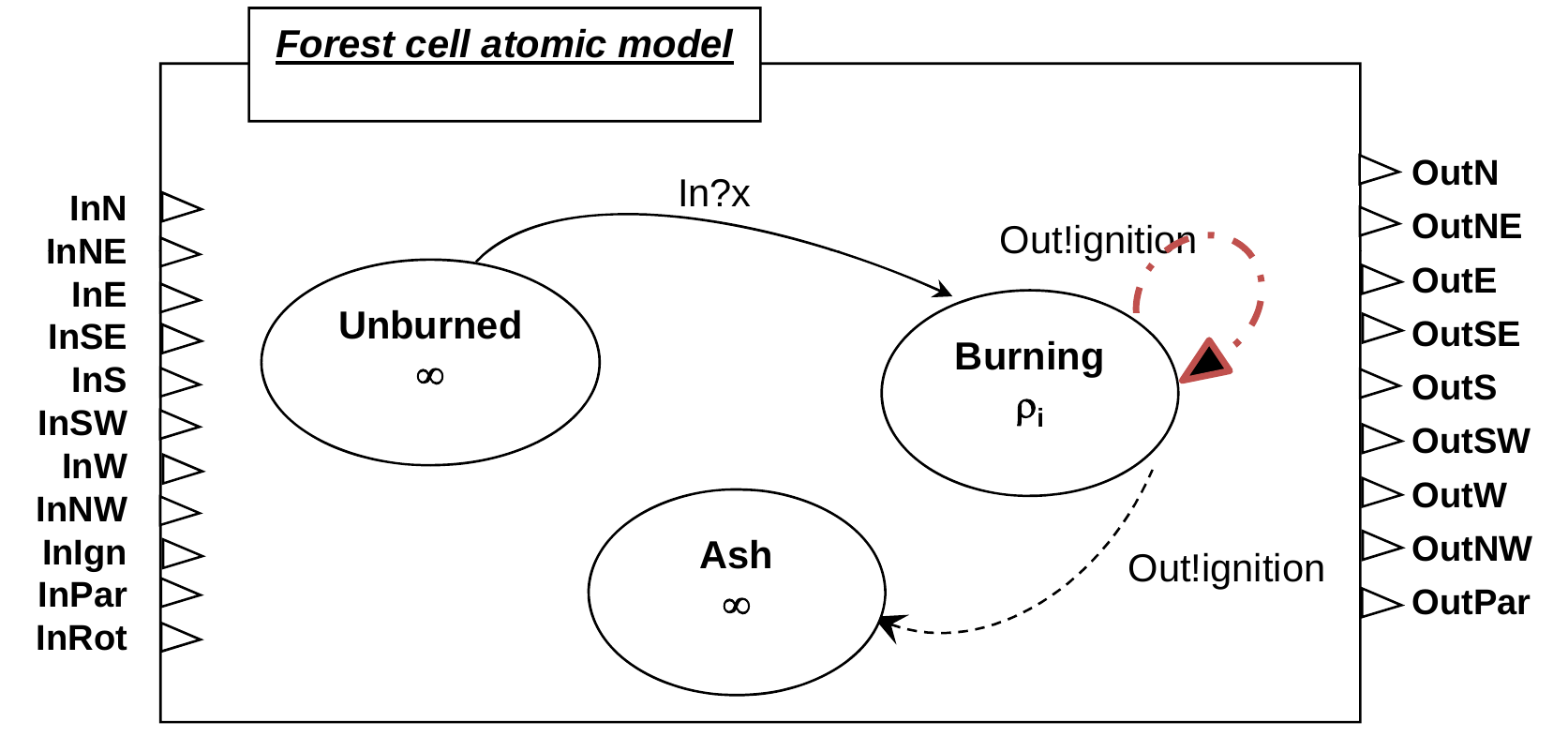}}
%.49\linewidth
\caption{Forest cell DEVS behavior.}
\label{fig:forestcelldevsbehavior}
}
\end{figure}

\newpage
The formal specification of the forest cell DEVS atomic model (Figure~\ref{fig:forestcelldevsbehavior}) is:\\

$ForestCellAM  = (X, S, Y, \delta_{ext}, \delta_{int}, \delta_{conf}, \lambda, t_a)$ where:     

$InPorts=\{ InN, InNE, InE, InSE, InS, InSW, InW, InNW, InIgn, \\InPar, InRot\} $\\
$XInPorts={R}^{9}\times{R}^{d}\times {R}^{2}$\\
Where dimension 9 represents the nine ignition messages (8 from neighbors and one from igniter model), $d$ is the number of parameters of each forest cell (weather, wind, humidity, etc.), and the number 2 is the couple, rate of fire spread and its main direction.

$OutPorts=\{OutN, OutNE, OutE, OutSE,  OutS, OutSW, OutW, OutNW,\\ OutPar\}$ \\
$YOutPorts = {R}^{8}\times {R}^{d}$

We have 8 output ignition messages coupled respectively to the 8 neighbor cells; and $d$ is a dimension, that represents the cell parameters, its output is coupled with Rothermel model. 

$X = \{(in, x)| in \in InPorts, x \in XInPorts\}$\\
$S = \{(Nonflammable,\infty),(Unburned,\infty),(Burning,\rho), (Ash, \infty)\} | \rho \in {R}^{+}$\\
$Y = \{(out, y)| out \in OutPorts, y \in YOutPorts\}$\\
$\delta_{int} (Burning, \rho) = (Burning, \rho_{i})\\
\hspace*{2cm} \hbox{if} (\Delta T_{td} \neq \emptyset )\hbox{ } \Delta T_{td} = \Delta T_{td} -{\rho}\\ \hspace*{4cm} \rho_{i}=min(\Delta T_{td})$\\
$\delta_{int} (Burning, \rho) = (Ash, \infty)\\
\hspace*{2cm} \hbox{if} |\Delta T_{td}| = 1 (\hbox{i.e. } \Delta T_{td} \hbox{ has a cardinality of 1}) $ \\
$\delta_{ext}(Unburned,  \rho,  (In ? x) ) = (Burning, \rho_{1})$\\ 
$\delta_{ext}( Burning,  \rho ,  (InPar ? x)  ) = (Burning, \rho_{i})$\\ 
$\delta_{conf}(s, \rho, x ) = \delta_{ext}(\delta_{int}(s),0,x)$\\
$\lambda(Burning) = Out ! ignition$\\
$t_{a}(phase, \rho)  = \rho$\vspace{0.5cm}\\
The initial state of this model is $(Unburned, \infty)$.\\
Where $In \in InPorts-\{InPar, InRot\}, \rho^{'}_{i}$ is computed by the Rothermel model whenever parameters change, $ Out \in OutPorts-\{OutPar\}$, $ignition\in R$.\\

In Figure~\ref{fig:forestcelldevsbehavior}, the solid line describes the occurrence of external event, while the dashed line depicts the internal state changes, and occurs when the elapsed time reaches the lifetime of the state.

\subsubsection{Forest Fire Spread and Decomposition}
\label{ForFirSprDec}

\begin{figure}[b]
{
\centering
%\figure[The LSIS\_DME approach.]
{\includegraphics[trim= 0mm 110mm 108mm 0mm, clip, height = 5 cm, width=0.8\linewidth]{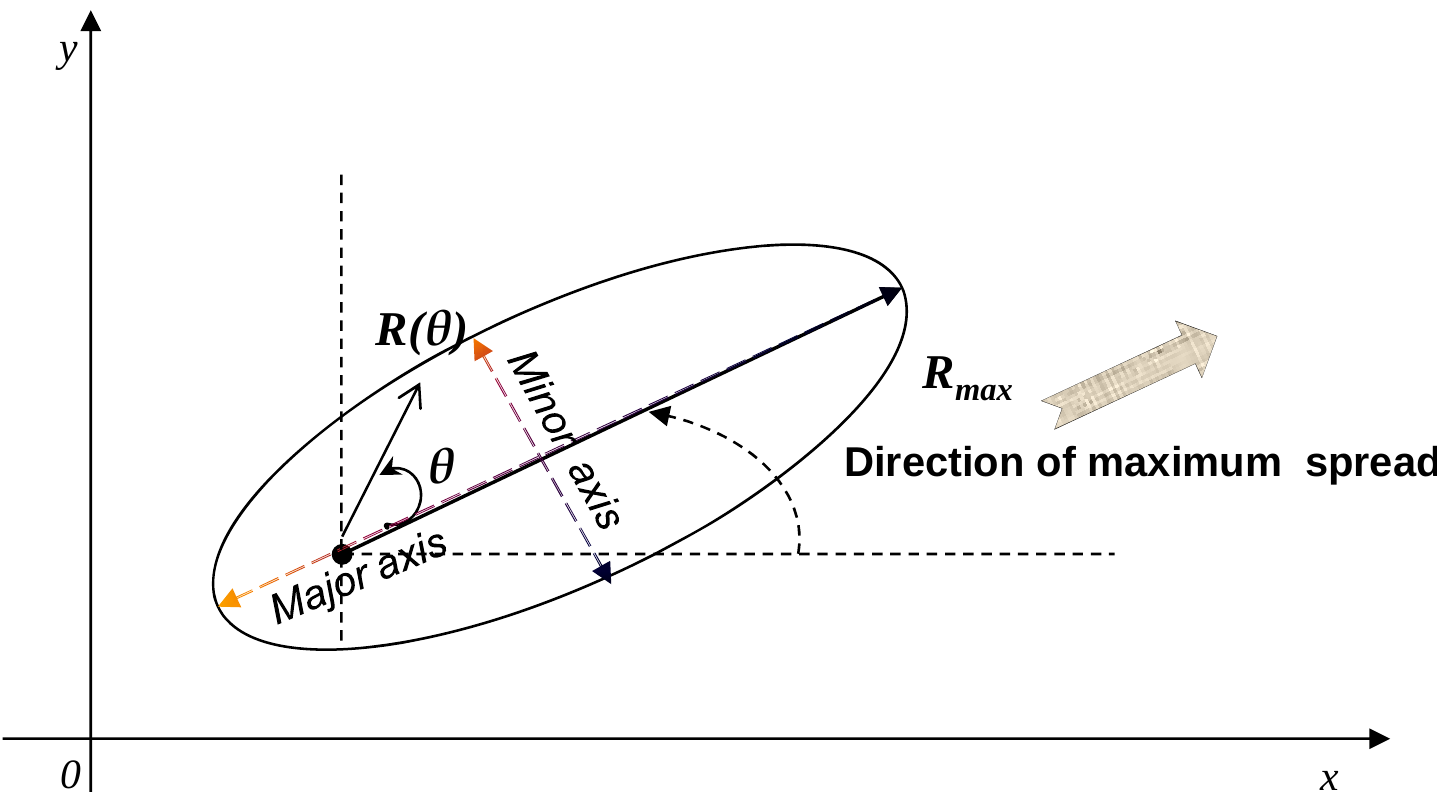}}
%.49\linewidth
\caption{Rate of spread according to an arbitrary direction.}
\label{fig:rosanydirec}
}
\end{figure}

The Rothermel mathematical model \cite{Rothermel1972} computes a one-dimensi- onal maximum fire spread rate and its direction. Fire geometry has been determined empirically \cite{Finney2004}. However, in the literature, the fire shape is considered as an ellipse which is widely used \cite{Anderson1983}. Therefore the fire spread is inferred in all directions from the 1-D maximum rate using the mathematical properties of the ellipse and the equations defined in \cite{Finney2004, DAmbrosio2006}. The spread rate in an arbitrary direction $\theta$ (Figure~\ref{fig:rosanydirec}) is obtained by:
\begin{eqnarray}
\label{eq:ros1}
R(\theta) & =  & R_{max} \frac{1-\epsilon}{1-\epsilon\cos \theta}\\
\label{eq:ros2}
\epsilon & = & \frac{\sqrt{lw^{2}-1}}{lw} \\
lw & = & 0.936e^{0.2566v} + 0.461e^{-0.1548v} - 0.397
\label{eq:ros3}
\end{eqnarray}

Where $R_{max}$ is maximum rate of spread, $lw$ is the ellipse ratio of the semi-major over semi-minor (Length to breath ratio) and $v$ is the midflame wind speed (effective wind speed).

Once the rate of spread is known in all directions, we compute the time delays to reach the center of the neighbors by this equation: 
\begin{eqnarray}
t_{i} & = &  \frac{d_{i}}{R_{i}} | i = 1..8
\label{eq:rosdirec}
\end{eqnarray}

Where $t_{i}$ is the time to reach the neighbor $i, d_{i}$ is the distance from the center of the burning cell to the center of its neighbor $i$ and $R_{i}$ is the fire spread rate in this direction.
In case of parameters change (wind speed, direction, humidity, etc.), the new  $R^{new}_{i}$ in direction of the neighbor $i$, is inferred from Rothermel model if the fire hasn't yet reached it. The new remaining time delay is:
\begin{eqnarray}
t^{new}_{i} & =& \frac{\Delta d_{i}}{R^{new}_{i}}
\label{eq:rosdirection}
\end{eqnarray}

Where $\Delta d_{i}$ is the remaining distance to reach the neighbor $i$ and $t^{new}_{i}$  is the new time delay.
As the fire propagates, the equation~\ref{eq:rosdirection} allows the model to be dynamic and sensitive to weather changes.

\subsubsection{System Formal Design by Closure under Coupling}
\label{sec:formaldesignCuC}
\begin{figure*}[!h]{
\setlength{\fboxsep}{0pt}%
\setlength{\fboxrule}{0pt}
\center
{\includegraphics[trim= 0mm 30mm 30mm 0mm, clip, height = 11 cm, width= 11cm]{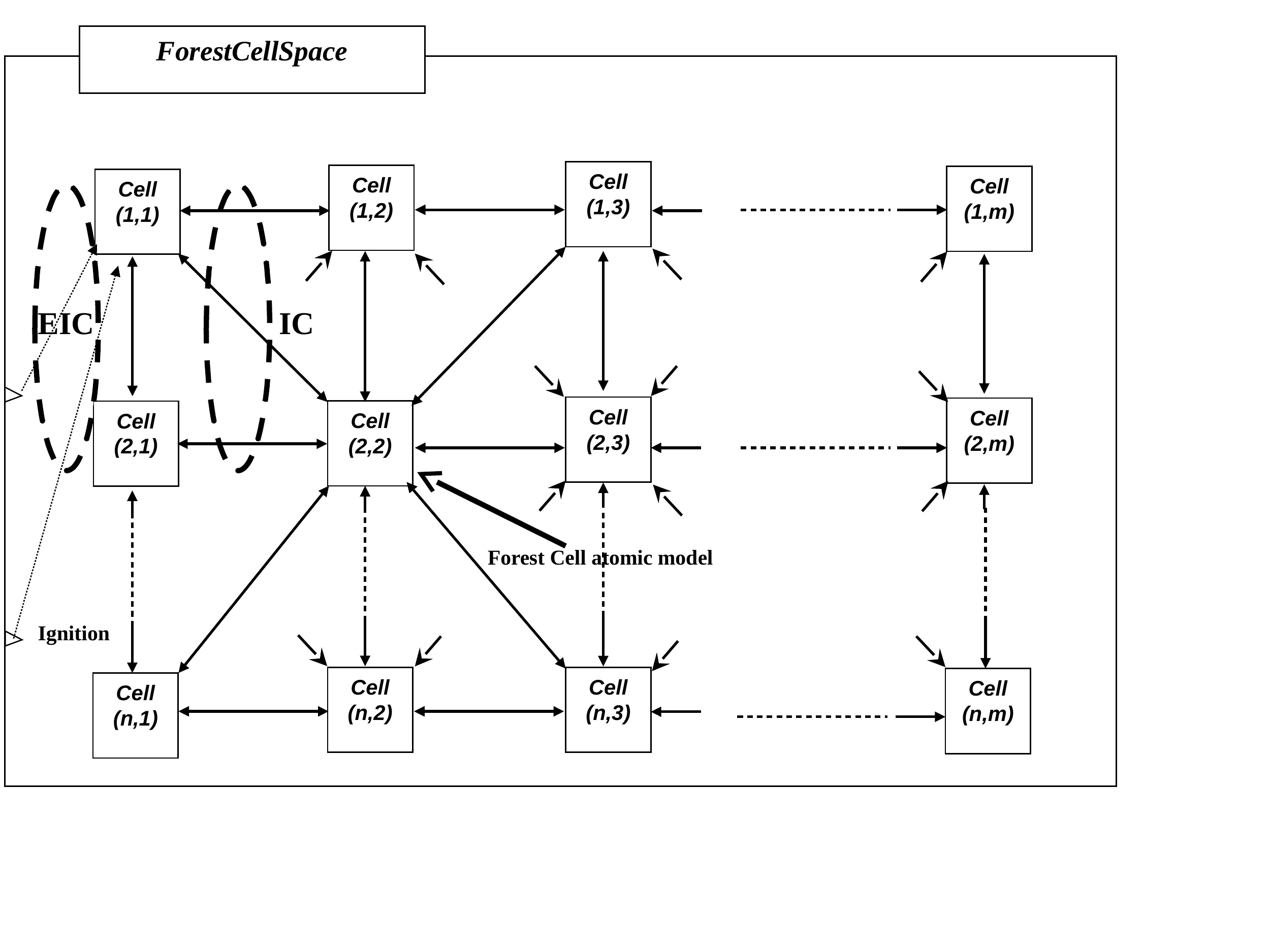}}
%.49\linewidth
\caption{Forest cell space DEVS coupled model.}
\label{fig:forestspread}
}
\end{figure*}

In the forest cellular approach, the forest fire spread is usually presented as a set of arranged cells whose dimensions depend on the resolution of GIS \cite{Hu2012}. Thus, forest cell space coupled model is a discrete dynamical system formed by coupling a finite number of forest cells. These cells are arranged uniformly in a two-dimensional space composed of $n$ rows and $m$ columns. Each forest cell atomic model is coupled to 8 neighbors as described in Figure~\ref{fig:forestspread}. 

The fire spread model is described in Section 4.2. It is composed of forest cell space (a grid of forest cells), igniter model, GIS interface and Rothermel model. However, simulate such a model will be affected by large-scale forest cell model and particularly during the simulation run and involves more memory usage. A huge memory is needed to simulate the overall forest and if you lack memory, you simulate just a part of it. In \cite{Ntaimo2008}, memory required for some implementations is illustrated. 

To overcome to this issue, we propose to use the closure under coupling. This property of the DEVS formalism consists to transform a coupled model into its equivalent atomic model. Thus, the forest cell space coupled model will be converted into an atomic model.

Each cell becomes a state variable of this atomic model and accedes to its neighbor state directly by removing the ports from the atomic models that is resulted by small volume of inter-component messages during simulation run. In fact, the communication between cells is done directly inside the model, except outdoor events~\cite{Shiginah2011}.

Based on the Forest Cell DEVS atomic model described in Section 4.3.1, the forest cell space DEVS coupled model is transformed into forest cell space DEVS atomic model and would benefit from the property of closure under coupling. It is described as follows:

$ForestCellSpaceAM = (X,S,Y, \delta_{int},\delta_{ext},  \delta_{conf}, \lambda, t_{a})$

\begin{itemize}

\item $X = \{(ignite, list), (inParam, parameters)\}$ | list is a list of initial ignited cells 
\item $Y = \emptyset$, there is no output event to send out.
\item $S = Cell\times inParam =\{\ldots…,(phase_{i,j},\rho_{i,j}),\ldots…\} \times {R}^{d} | i=1..n \hbox{ } j=1..m$

\end{itemize}
Where \textit{Cell} is the set of forest cell. Each cell is identified by its position ($row$, $column$) which can be in one of the four states (Nonflammable, Unburned, Burning, Ash) and $d$ is the number of parameters provided by GIS interface.
\\
Each cell keeps its current state and according to external and internal events state changes will occur to update the state of the concerning cell. It is pointed out that each cell has eight neighbors except those situated on the bound of the forest. Environmental parameters are assumed to be uniform at each cell except the wind which can change its speed and direction over time. Thus, the fire spread direction can change dynamically. 
\begin{figure}[!h]
%\vspace{-0.2cm}
\small{
\hspace*{0.5cm}{$\bf\delta_{ext}(s, e, x)$}\\
\hspace*{1.5cm}{c, c\'{}  : Cell}\\
\hspace*{1.5cm}{if (x = ignite)\{}\\
\hspace*{1.75cm}{for each c $\in$ ignite$_{list}$ }\\
\hspace*{2cm}{s$_{c}$ = $\delta_{ext}$ (s$_{c}$, e, ignite)\}}\\
%\hspace*{1.5cm}{\}}\\
%{recompute the lifetime for each active cell $c$}\\
\hspace*{1.5cm}{if (x = inParam)}\\
\hspace*{1.75cm}{update parameters}\\
\hspace*{1.5cm}{recompute the lifetime for each active cell $c$}\\
%%delta_int
\hspace*{0.5cm}{$\bf\delta_{int}(s)$ }\\
\hspace*{1.5cm}{for each $c \in Cell$\{}\\
\hspace*{1.75cm}{if (lifetime(c) = lifetime(Cell))\{}\\
\hspace*{2cm}{if (burning($s_{c}$))}\\
\hspace*{2.25cm}{for each c\'{} $\in$ neighbor(c)}\\
\hspace*{2.5cm}{ s$_{c'}$ = $\delta_{ext}$(s$_{c'}$, lifetime(Cell), ignite)}\\ 
\hspace*{1.75cm}{$s_{c} = \delta_{int}(s_c)\}$}\\
%\hspace*{1.75cm}{\}}\\
\hspace*{1.75cm}{else }\\
\hspace*{2cm}{ lifetime(c) = lifetime(c) - lifetime(Cell) \}}\\
%\hspace*{1.5cm}{\}}\\
\hspace*{0.5cm}{$\bf \delta_{conf}(s, e, x)$ }\\
\hspace*{1.5cm}{$\delta_{ext}(\delta_{int}(s), 0, x)$}\\
\hspace*{0.5cm}{$\bf lifetime(s)$ }\\
\hspace*{1.5cm}{return min \{lifetime(c) $\vert$ $c \in Cell$\}}\\
}
\caption{DEVS atomic model functions of forest fire spread}
\label{fig:devsmodelfunctions}
\end{figure}
Thus, we get a non-modular modeling structure (see Figure~\ref{fig:devsmodelfunctions}). With a modular modeling, when a cell ignites its neighbors, it invokes internal state change to provoke the execution of the output function $\lambda()$, which sends out the event ignition to all neighbors. On the other hand, the non-modular modeling propagates the output event ignition directly to neighboring cells without calling the simulator for dispatching events. Such a communication optimizes the DEVS M\&S structure and decreases the number of exchanged messages between cells.

%In  Figure~\ref{fig:devsmodelfunctions} the transition functions of our model.

\begin{figure}[!h]
{
\centering
%\figure[The LSIS\_DME approach.]
{\includegraphics[trim= 0mm 110mm 90mm 0mm, clip, height = 5.5 cm, width=8.5cm]{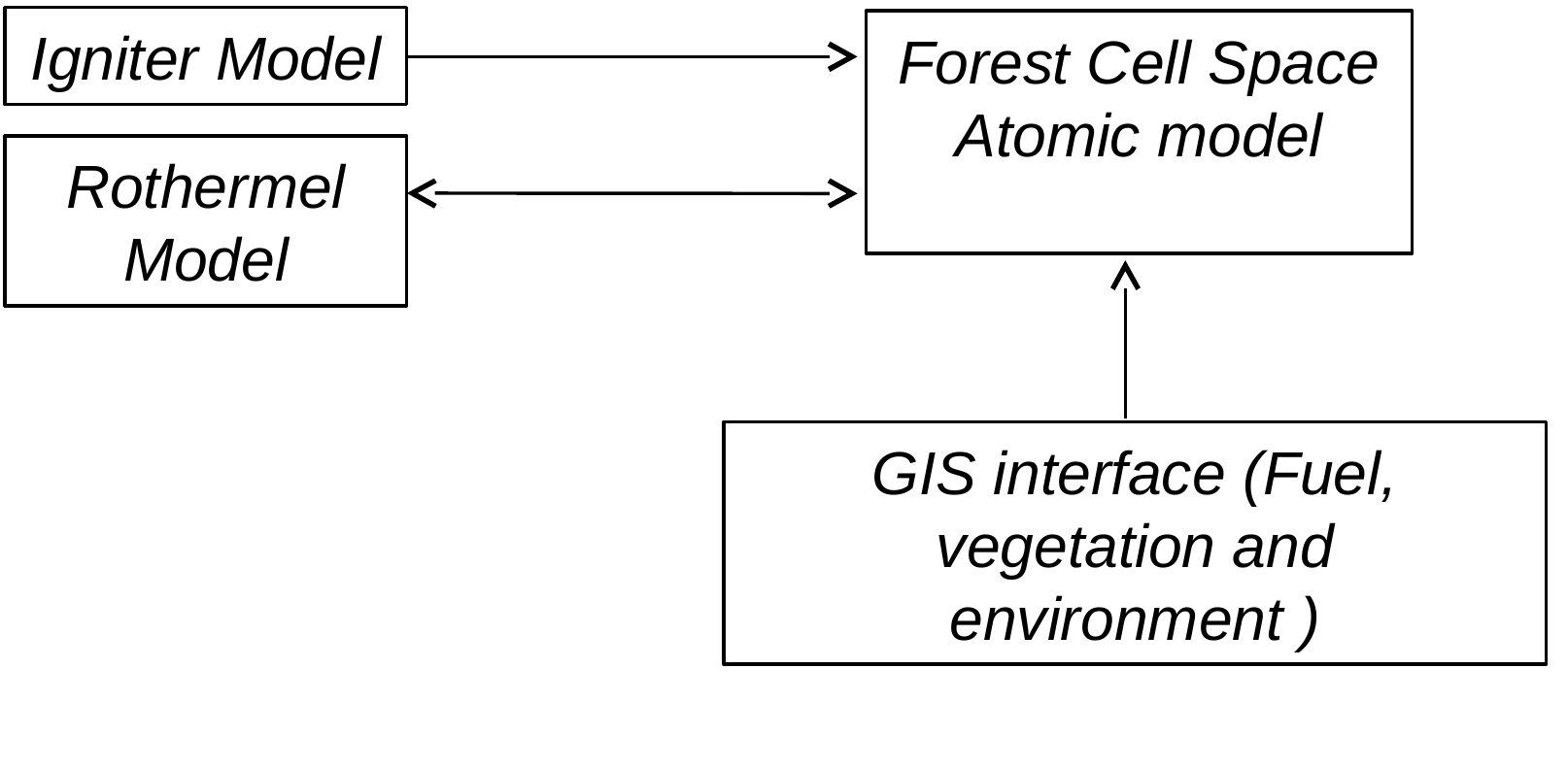}}
%.49\linewidth
\caption{The forest fire spread model via closure under coupling.}
\label{fig:newfuncarchi}
}
\end{figure}

Consequently, the fire spread model in its non-modular structure is comprised of four atomic models: Igniter atomic model, Forest cell space atomic model, GIS interface and Rothermel model as shown on Figure~\ref{fig:newfuncarchi}. It is identical to the structure depicted in Figure~\ref{fig:funcarchisim} except the Forest Cell Space which is converted to an atomic model.

Note that this architecture remains an open and modular one which allows enhancing the functionalities of the GIS and updating the Rothermel package.
\\
\\
\section{Simulation Experiments}
\label{sec:experiments}
In this section some output results that measure the simulation performance of the two models are provided. The package Behave was used to compute the rate of fire spread given by the Rothermel model.  The experiments were carried out on laptop Dell T2300@1.66GHz with Intel CPU, 1.66 GHz processor, 2.50 GB of RAM and Windows XP\copyright{}  operating system and as a simulator, we have used our own simulator developed in Java and respecting DEVS specifications.     

\subsection{Comparison Results}

As mentioned by \cite{Ntaimo2008}, an assessment of memory usage based on the implementation in  \cite{DEVSJAVA2012} shows that each cell needs about 35 kB memory space. This fact is time consuming and resource intensive and therefore not very effective for large-scale cells simulation.    

We were tempted to see the limits of our simulator for these two specifications of forest fire spread. The ForestCellSpace DEVS coupled model versus ForestCellSpace DEVS atomic model.

Table~\ref{tab:comparison} summarizes some important results. The model ForestCellSpace coupled model uses a conventional architecture, where a grid of cells is used. While ForestCellSpace atomic model uses the closure under coupling to get a unique atomic model of the forest cell space.

\begin{table}
\centering
\renewcommand{\arraystretch}{1}
\caption{Comparison results.}
\small{
\centering
\begin{tabular}{p{5cm}p{2.5cm}p{2.5cm}}
\hline
{\bf  Results}  & {\bf  ForestCellSpace Coupled Model} & {\bf  ForestCellSpace Atomic Model} \\
\hline
Maximum number of simulated cells [cells] & 70$\times$70 & 2682$\times$2682 \\
\hline
Used Memory [MegaBytes (MB)]  for 60$\times$60 cells  & 8.74 & 1.23 \\
\hline
Used Memory [MB] for 600$\times$600 cells & Out of Memory Error & 12 \\
\hline
Simulation Execution Time  [second (sec)] for 60$\times$60 cells & 109,131 & 13,390 \\
\hline
\end{tabular}
}
\label{tab:comparison}
\end{table}

The experiment revealed that only about 70$\times$70 cells of the overall forest space has been simulated in coupled model, whereas the other specification, i.e., forest cell atomic model, a limit of 2682$\times$2682 cells was reached  therefore, we get a ratio close of 1 over 39 for each cell.

Another advantage of the forest cell space atomic model is the simulation time; in fact, the latter is 8.15 times faster than the forest cell space coupled model.

\begin{figure}
{
\centering
%\vspace{-0.1cm}
\subfigure[ForestCellSpace coupled model: grid of cells.]{
\centering
\includegraphics[trim= 0mm 1mm 1mm 0mm, clip, height = .3\paperheight, width= \linewidth ]{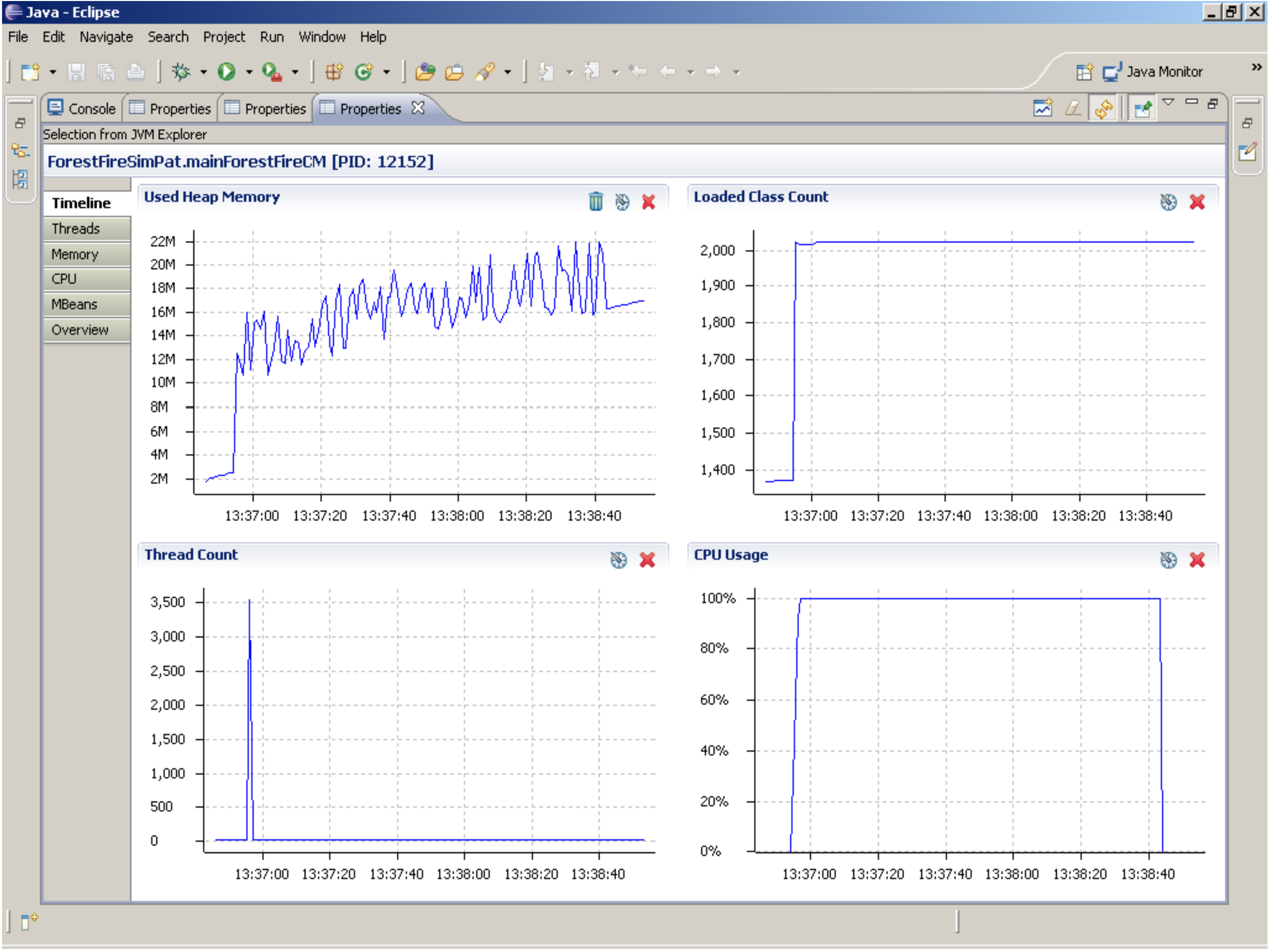}
\label{fig:DEVScoupledsim}
}
%\vspace{-0.1cm}
\subfigure[ForestCellSpace atomic model: closure under coupling.]{
\centering
\includegraphics[trim= 0mm 1mm 1mm 0mm, clip, height = .3\paperheight,  width= \linewidth ]{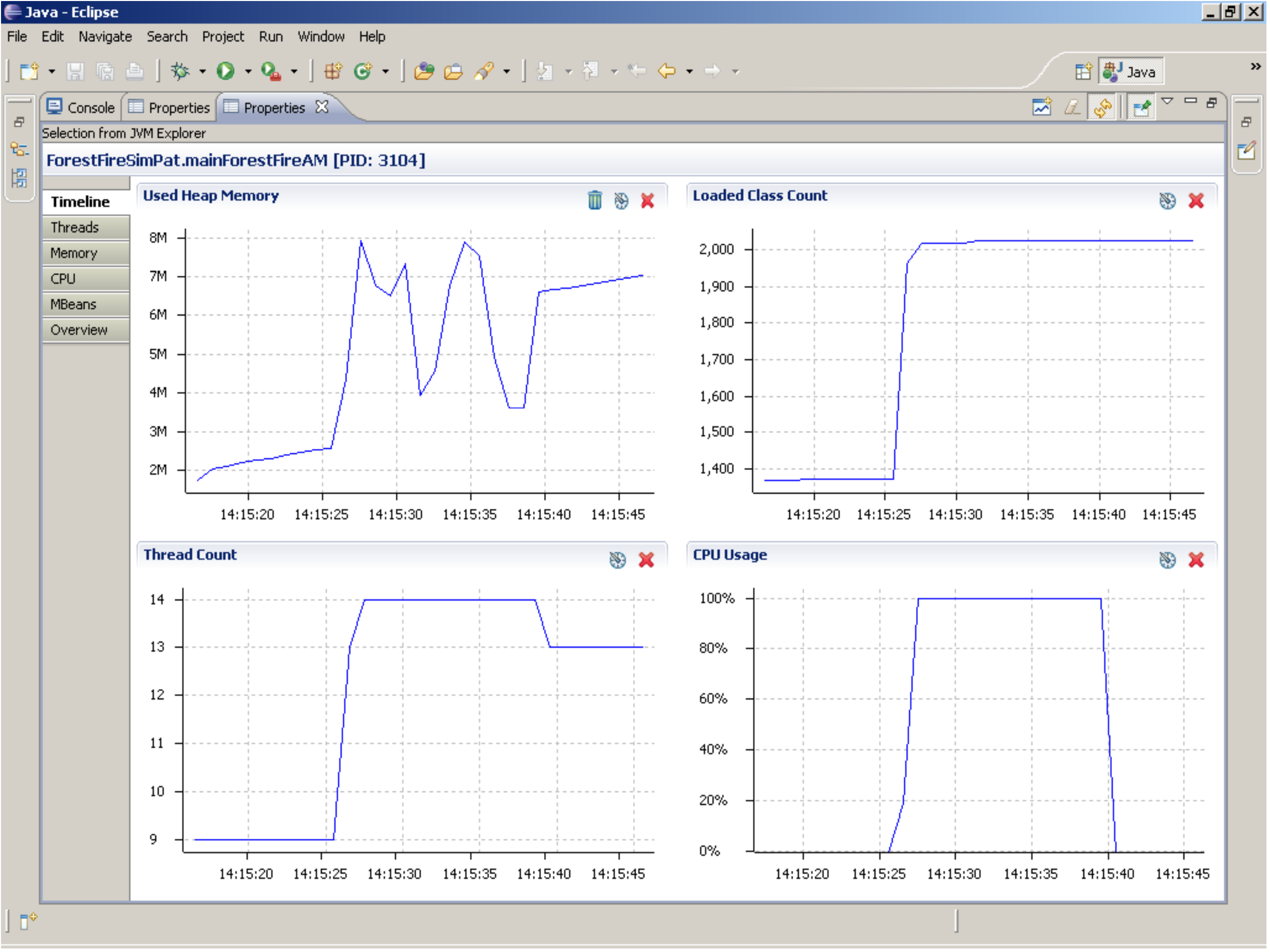}
\label{fig:DEVSatomicsim}
}
\caption{Performances of forest fire spread simulation.}
\label{fig:perfofforest}
}
\end{figure}

In order to get more results, we have used additional free software Java Virtual Machine Monitor (JVM Monitor) to profile our Java application. The snapshot of the Figure~\ref{fig:DEVScoupledsim} resumes some characteristics of simulation of ForestCellSpace coupled model; whereas Figure~\ref{fig:DEVSatomicsim} gives the values of these characteristics of ForestCellSpace atomic model. 

Through these Figures~\ref{fig:DEVScoupledsim}~\&~\ref{fig:DEVSatomicsim} and by looking inside execution time windows of each simulation, we can see the advantages of the usage of closure under coupling property. We have less amount of used heap memory. The maximum values reached with Forest Cell Space Coupled Model are greater in overall cases compared to Forest Cell Space Atomic Model. 

Table~2 gives the performance indicators of both Forest Cell Space simulations coupled and atomic at some critical times which are extracted from curves of Figure~\ref{fig:perfofforest} in case of $60\times60$ cells.

\begin{table}
\renewcommand{\arraystretch}{1}
\centering
\small{
\caption{Sample of simulation performances of ForestCellSpace:  Coupled vs. Atomic}
\centering
\begin{tabular}{p{4cm}p{2.5cm}p{2.5cm}}
\hline
{\bf Indicator} & {\bf ForestCellSpace Coupled Model} & {\bf ForestCellSpace Atomic Model} \\
\hline
Simulation Execution Time [sec.] & 109,131 & 13,390\\
\hline
Used Heap Memory [MB] & [2, 20] & [2, 8]\\
\hline
Thread Count & [9, 3500] & [9, 14]\\
\hline
CPU Usage [\%] & 100 & 100 \\
\hline
\end{tabular}
}
\label{tab:Tperfofforest2}
\end{table}

\begin{figure*}[!h]
{
\centering
%\vspace{-0.05cm}
\subfigure[DEVS coupled: time = 900 sec.]{
\centering
\includegraphics[trim= 0mm 10mm 70mm 0mm, clip, height=2.5cm, width=0.23\linewidth]{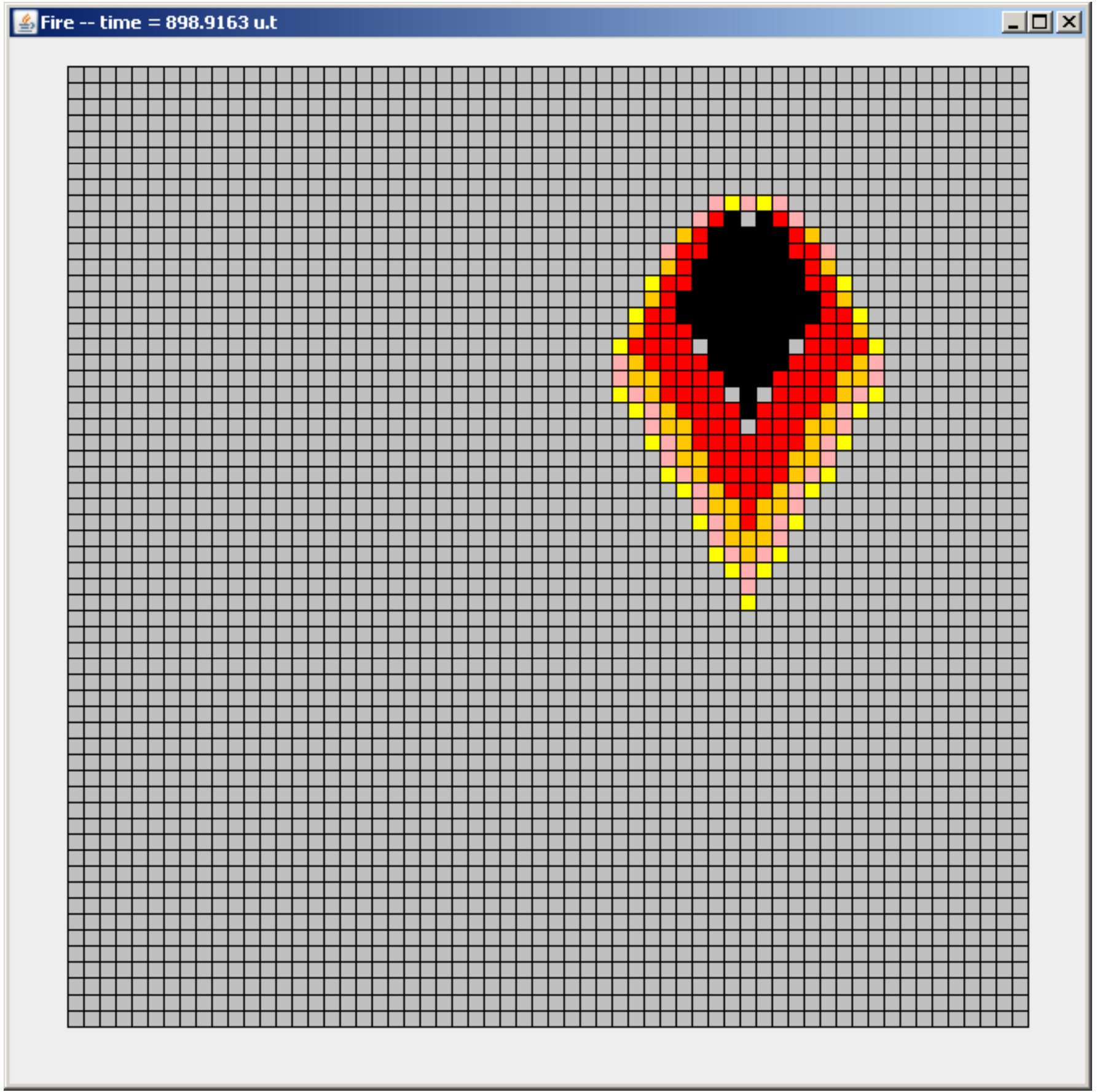}
\label{fig:DEVScoupled-1}
}
%\hspace{-0.1cm}
\subfigure[DEVS atomic: time = 900 sec.]{
\centering
\includegraphics[trim=  0mm 10mm 70mm 0mm, clip, height=2.5cm,width=0.23\linewidth]{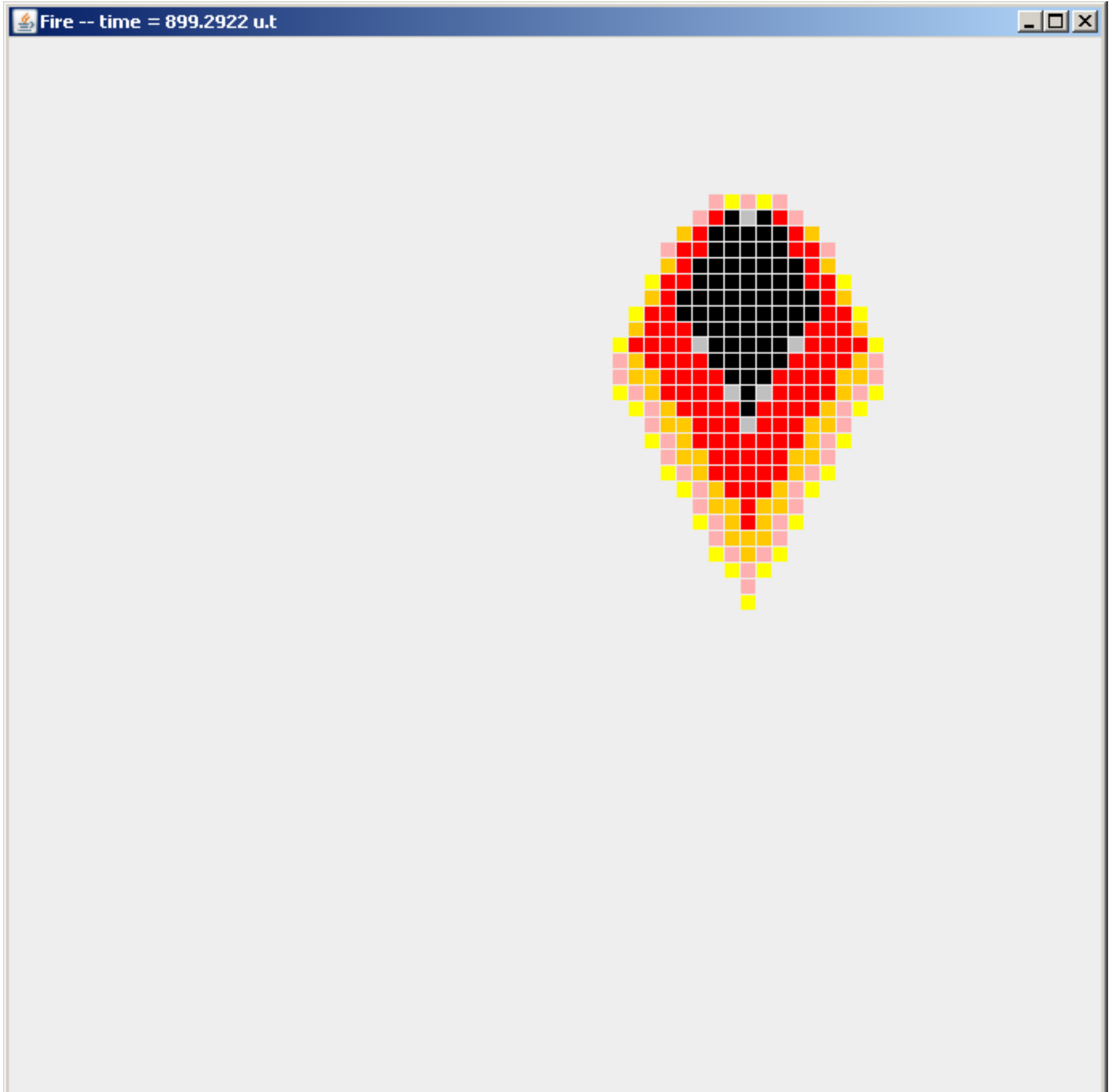}
\label{fig:DEVScoupled-2}
}
%\vspace{-0.05cm}
\subfigure[DEVS coupled: time = 1600 sec.]{
\centering
\includegraphics[trim=  0mm 10mm 70mm 0mm, clip, height=2.5cm,width=0.23\linewidth]{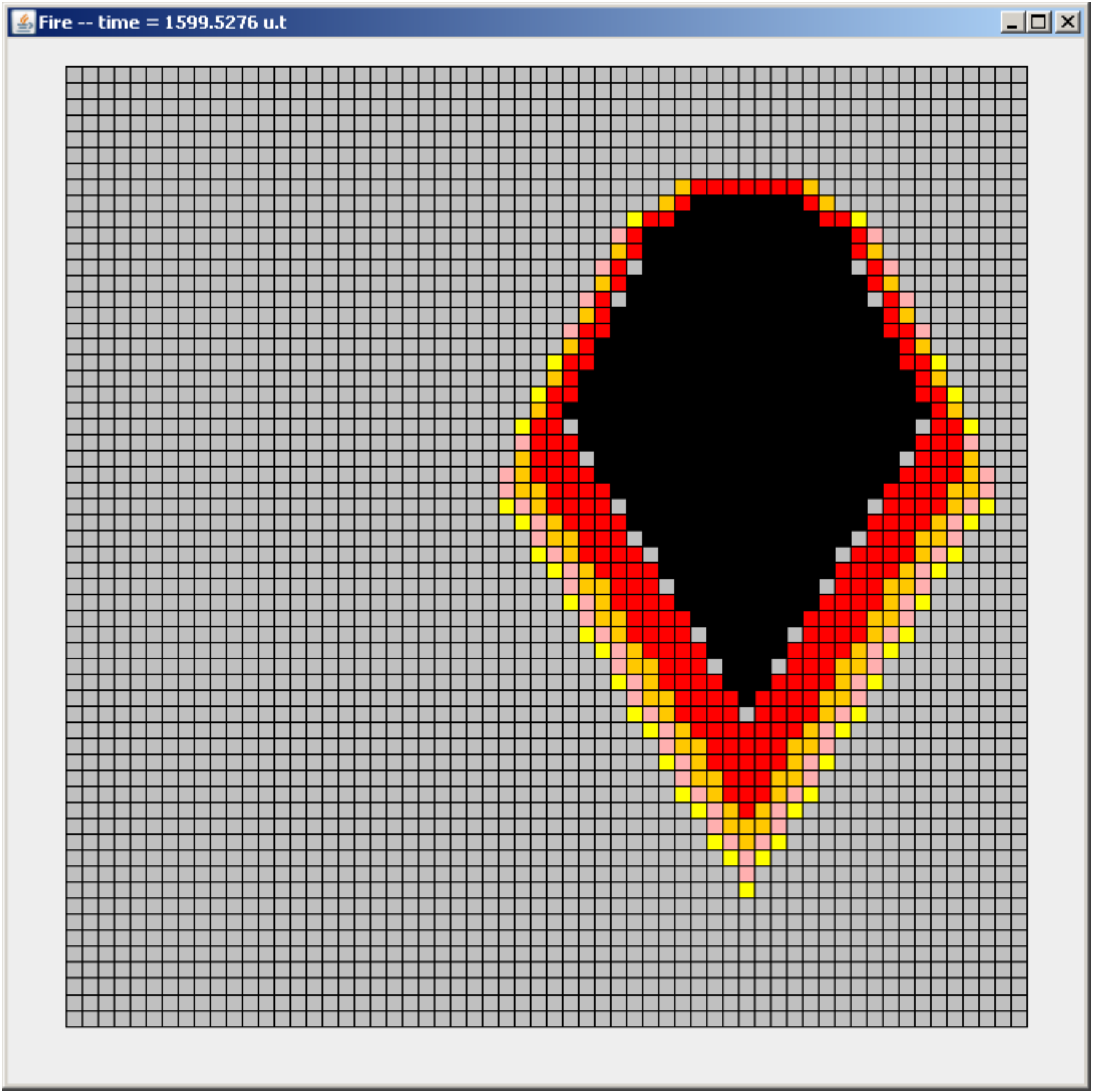}
\label{fig:DEVSatomic-3}
}
%\hspace{-0.1cm}
\subfigure[DEVS atomic: time =1600 sec.]{
\centering
\includegraphics[trim=  0mm 10mm 70mm 0mm, clip, height=2.5cm,width=0.23\linewidth]{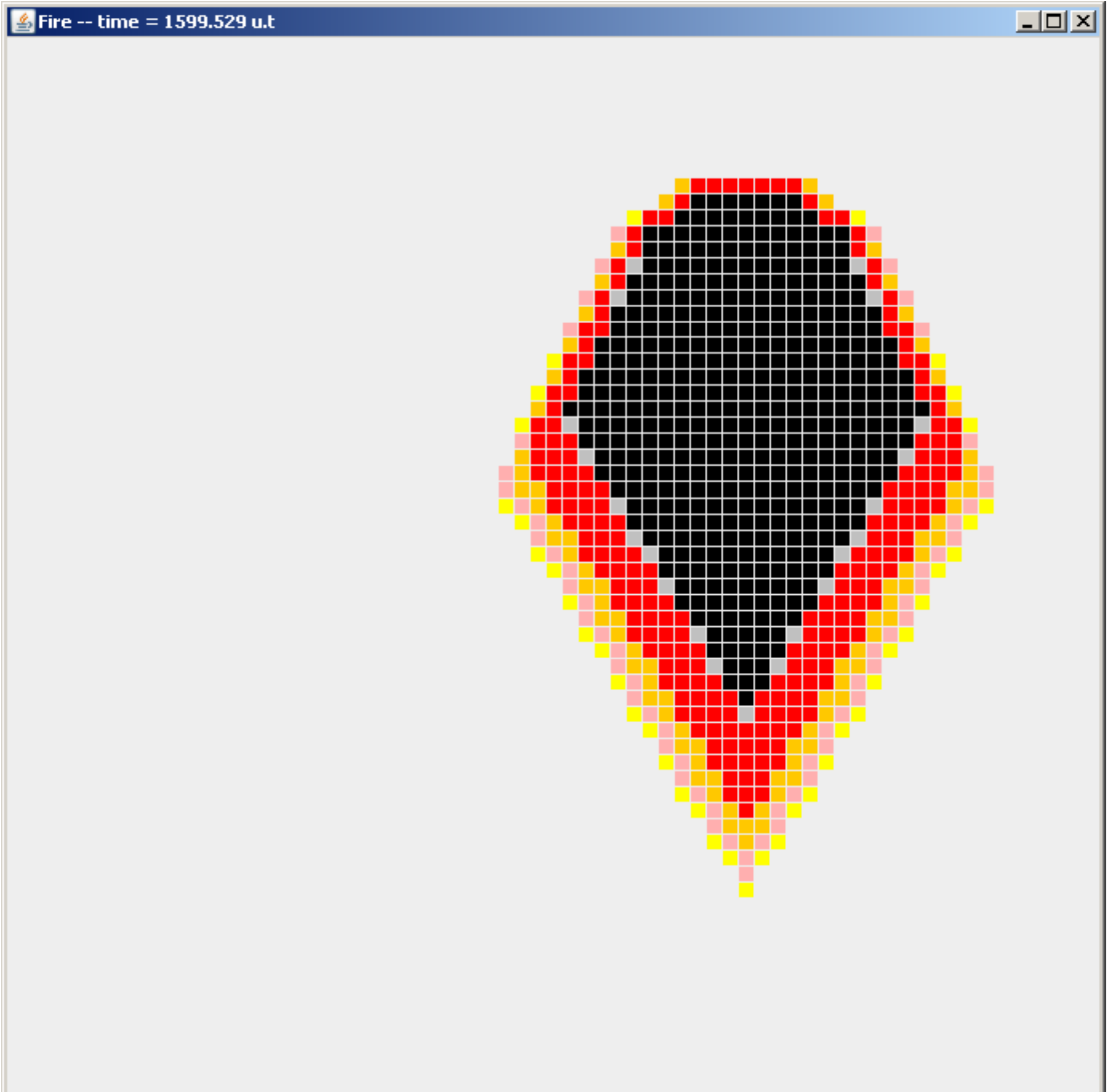}
\label{fig:DEVSatomic-2}
}
%\vspace{-0.25cm}
\caption{ForestCellSpace coupled model vs. ForestCellSpace atomic model.}
\label{fig:forestfire60x60}
}
\end{figure*}

As mentioned before, despite the different implementations of forest fire spread, it is essential to reminder that both models have the same behavior, whereas their simulation performances are different (see Figure~\ref{fig:forestfire60x60} for a cell space of 60$\times$60). 

The front flame is depicted in color yellow while the black color is dedicated to the ash area. The colors pink, orange, red and light gray represent the different phases of the burning state (flanking and backing spread).  

\subsection{Verification and validation}

The validation of fire spread model is not an easy task, and few works describe validations based on real fire \cite{DAmbrosio2006}. Nevertheless, we are going to do a partial validation by comparing the results of our approach by those of FARSITE. The latter is the most generally distributed and accepted fire behavior predictive model used in forestry \cite{Ntaimo2008}.

On one hand, we are going to simulate our model described in Section 4.2 by adopting  this functional architecture, on the other hand, we are going to simulate by FARSITE software the same scenario and compare the different results. Therefore, the equivalence of the two forms is verified using simulation methods.

In these simulations, we suppose that:
\begin{itemize}
\item Virtual forest is constructed as a grid of 260$\times$260 cells where each cell represents an area of 1$\times1$ m$^{2}$ (m: meter).
\item Fire spreading is on each of the eight compass points.
\item Starting ignited point is the cell(130,30).
\end{itemize}
We assume that uniform parameters characterize the cell space, i.e., each cell has uniform fuel, terrain, weather and fire behavior along the forest fire area. 

The fire shape is assumed elliptical and the decomposition scheme considered is center-to-center. We use the Rothermel model to compute the maximum fire spread rate. Fire spread in all other directions is inferred from the spread rate using the mathematical properties of the ellipse (equation~\ref{eq:ros1}).

We consider the following values to get the rate of spread by the Rothermel model:

The model of fuel considered in this experiment is Northern Fire Forest Laboratory (NFFL) Fuel Model 1: short grass (1 feet (ft)) with its standard characteristics.

Besides that, we assume:\\
\hspace*{1cm}$-$ wind speed [meter/second (m/s)] $wsp =  1.0$\\
\hspace*{1cm}$-$ wind direction [degree($^\circ$)] $wdr =      0.0$\\
\hspace*{1cm}$-$ slope [$^\circ$] $slp=  16.7  $\\
\hspace*{1cm}$-$ aspect [$^\circ$] $asp = 0.0$	\\		
The Rothermel model produces the following results:\\
\hspace*{1cm}  $-$ rate of spread  [m/s]   $ros = 0.26$\\
\hspace*{1cm}  $-$ direction of maximum spread  [$^\circ$] $ sdr  = 180$\\
\hspace*{1cm}  $-$ effective wind speed [m/s]     \textit{efw} $ = 1.37$\\

Length to breath ratio $lw$ is obtained via equation~\ref{eq:ros3}, its value for theses parameters is:\\
\hspace*{1cm}{$-$ $lw = 1.50$}

The spreading rate in each direction is given by equation~\ref{eq:ros1} based on the ratio $lw$. Thus, we obtain the following spreading rates according to each direction as shown on Table~\ref{tab:rosinputs}.

\begin{table}
\renewcommand{\arraystretch}{1}
\caption{Spreading rate in the eight directions}
\centering
\small{
\begin{tabular}{lll}
\hline
{\bf  $\theta$}  & {\bf  $R(\theta$)} & {\bf  \% $ros$} \\
\hline
0 & 0.26 & 100\\
\hline
$\mp$ 45 & 0.14 & 53.77\\
\hline
$\mp$ 90 & 0.066 & 25.41\\
\hline
$\mp$ 135 & 0.04 & 16.63\\
\hline
180 & 0.038 & 14.55\\
\hline
\end{tabular}
}
\label{tab:rosinputs}
\end{table}

\begin{figure}
{
\centering
%\vspace{-0.1cm}
\subfigure[time = 300 sec.: 0.29 ha ]{
\centering
\includegraphics[trim= 0mm 20mm 70mm 0mm, clip, width=0.23\linewidth]{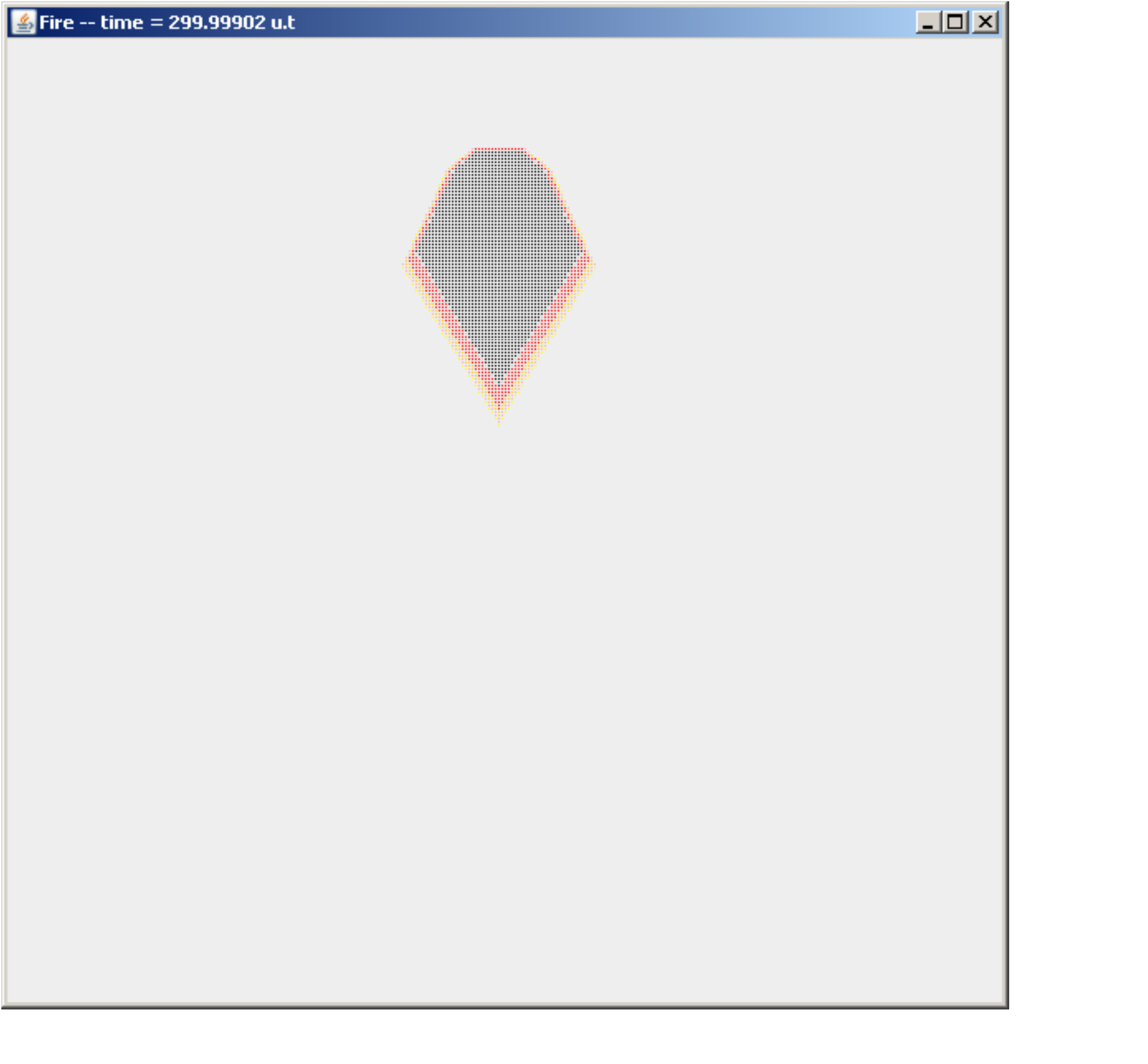}
\label{fig:DEVScoupled2-1}
}
%\vspace{-0.1cm}
\subfigure[time = 900 sec.: 1.17 ha]{
\centering
\includegraphics[trim= 0mm 20mm 70mm 0mm, clip, width=0.23\linewidth]{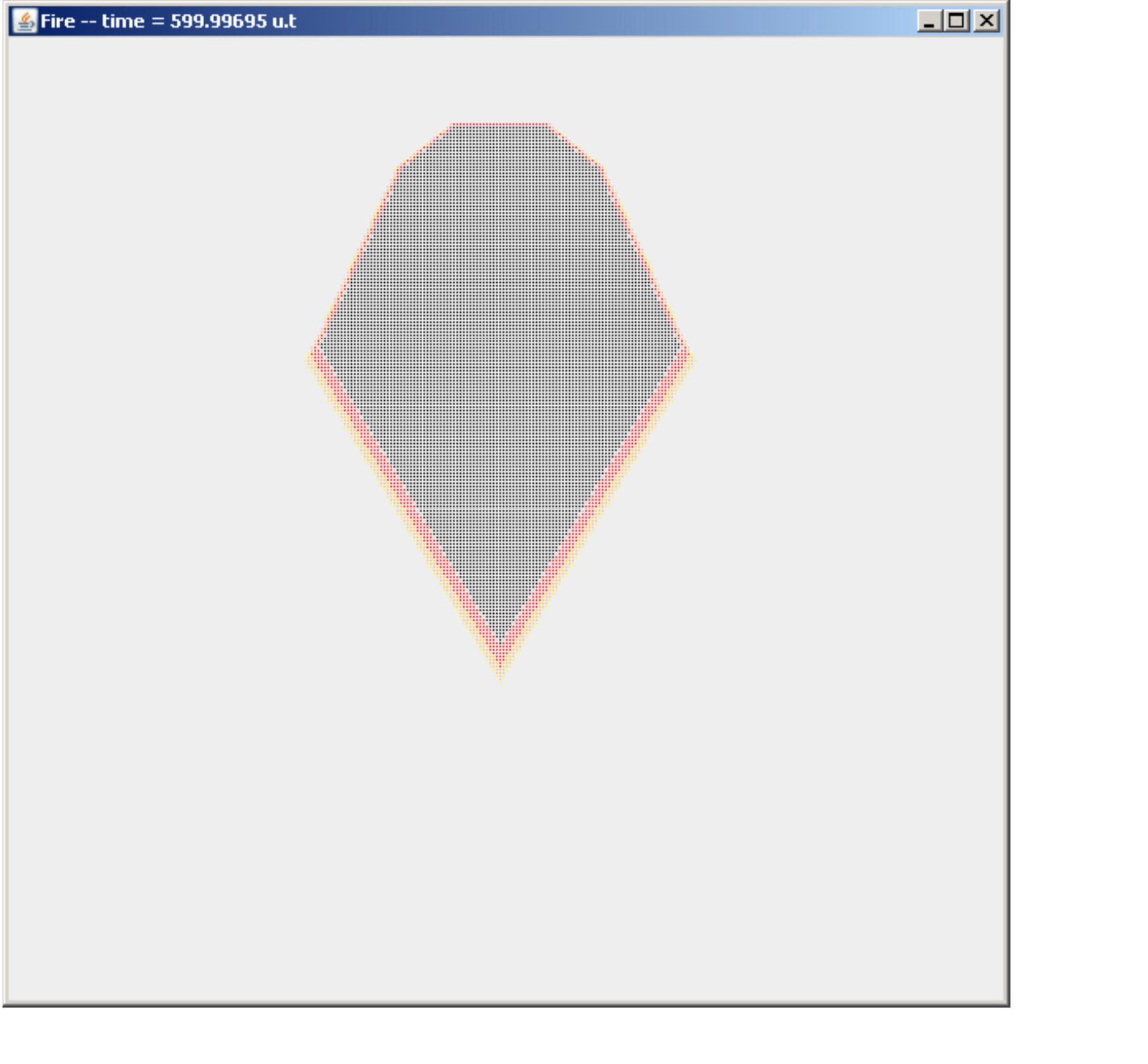}
\label{fig:DEVScoupled2-2}
}
%\vspace{-0.1cm}
\subfigure[time = 1200 sec.: 4.28 ha]{
\centering
\includegraphics[trim= 0mm 20mm 70mm 0mm, clip, width=0.23\linewidth]{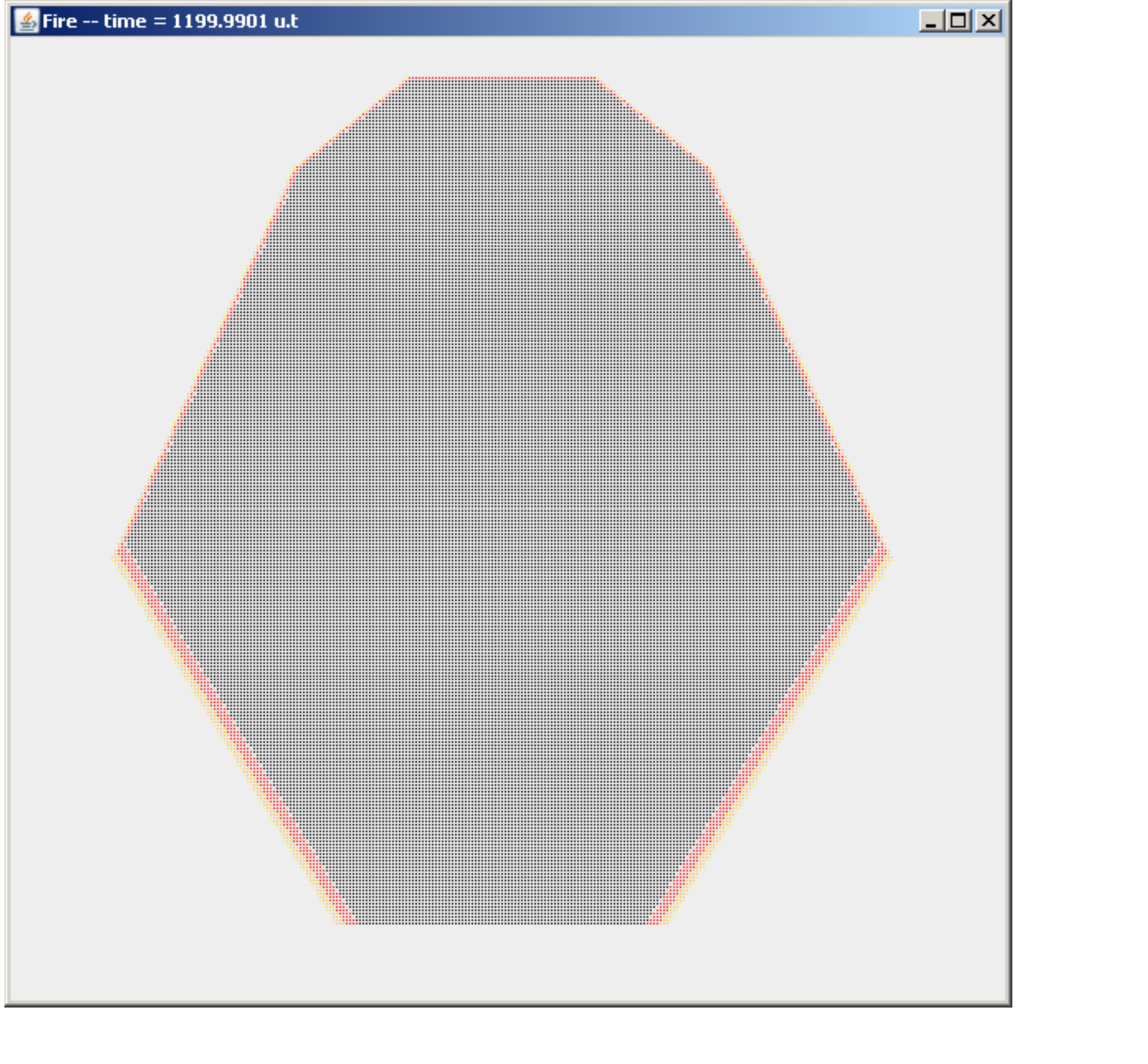}
\label{fig:DEVSatomic2-1}
}
%\vspace{-0.1cm}
\subfigure[time = 3184 sec.: 6.76 ha]{
\centering
\includegraphics[trim= 0mm 20mm 70mm 0mm, clip, width=0.23\linewidth]{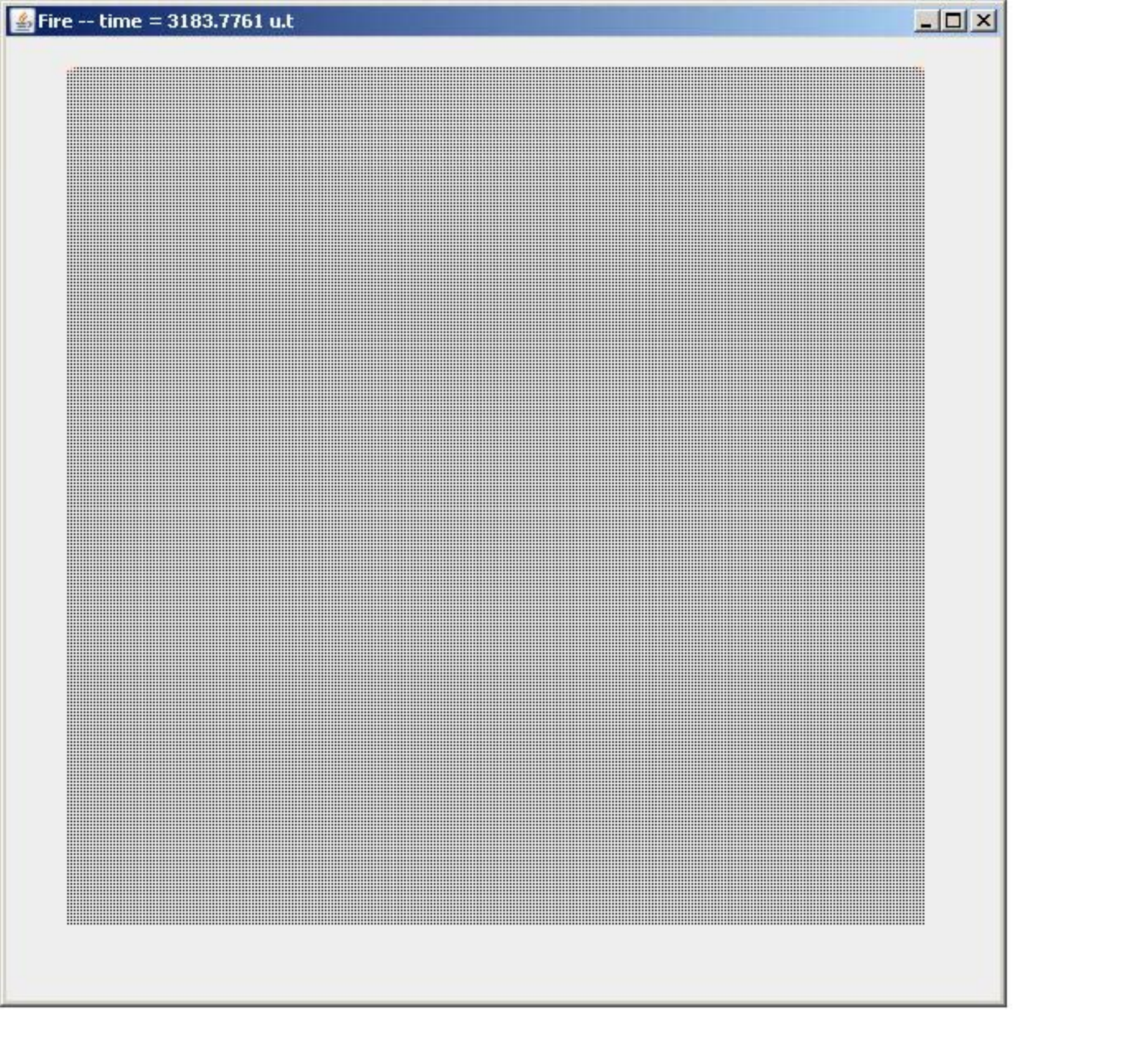}
\label{fig:DEVSatomic2-2}
}
\caption{Fire spread progression (260$\times$260 cells)}
\label{fig:forestcellsim260}
}
\end{figure}

According to these results, we simulate our forest fire model. Figure~\ref{fig:forestcellsim260} shows the fire spread taken at different simulation times and gives us the effective burned area [hectare (ha)]. 

\begin{figure}[!h]
{
\centering
%\vspace{-0.1cm}
%\begin{subfigure}
\subfigure[Fire perimeter.]{
\includegraphics[trim= 0mm 135mm 195mm 1mm, clip, height = 4.2cm,  width=0.47\linewidth]{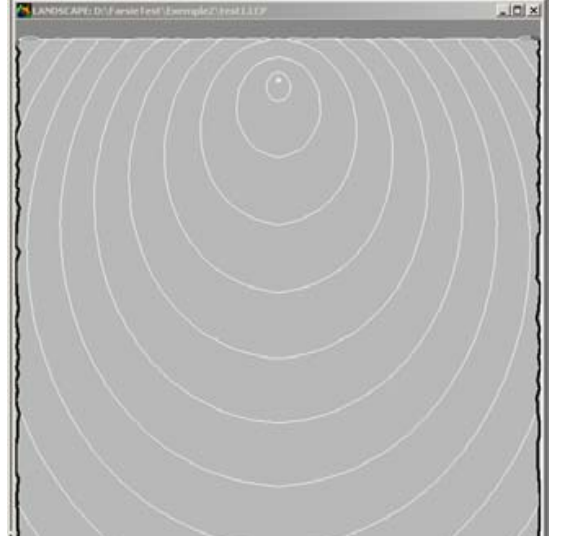}
\label{fig:farsite1}
}
%\vspace{-0.1cm}
\subfigure[Burned area curve.]{
\centering
\includegraphics[trim= 0mm 56mm 110mm 0mm, clip, height = 4.2cm, width=0.47\linewidth]{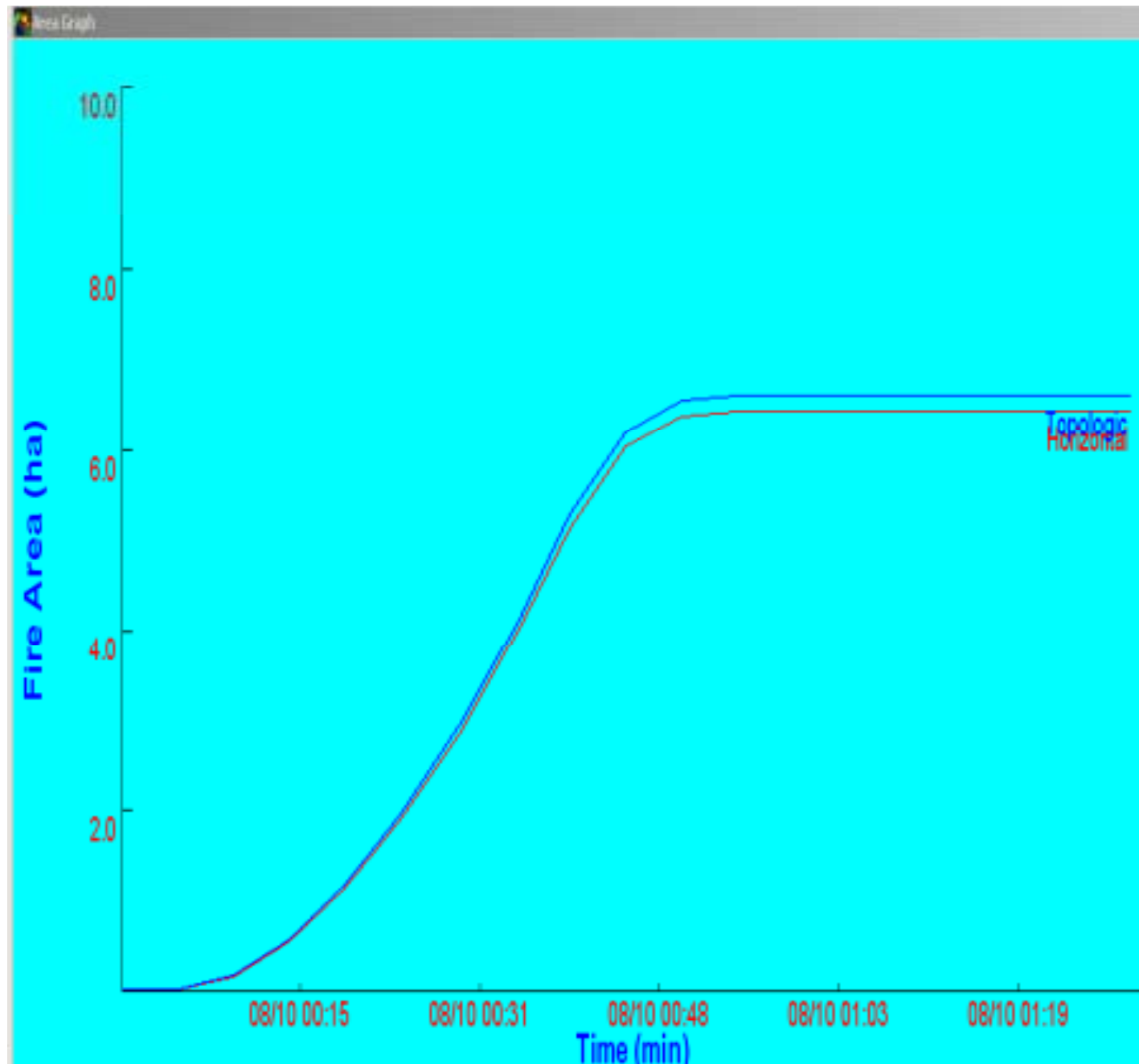}
\label{fig:farsite2}
}

\caption{Fire spread progression on FARSITE}
\label{fig:farsitesim}
}
\end{figure}

On the other hand, we have done the simulation on FARSITE  with the same parameters and we obtained output results shown on Figure~\ref{fig:farsitesim}.

\begin{figure}
{
\centering
%\figure[The LSIS\_DME approach.]
{\includegraphics[trim= 25mm 125mm 175mm 15mm, clip, height = 5cm, width=9cm ]{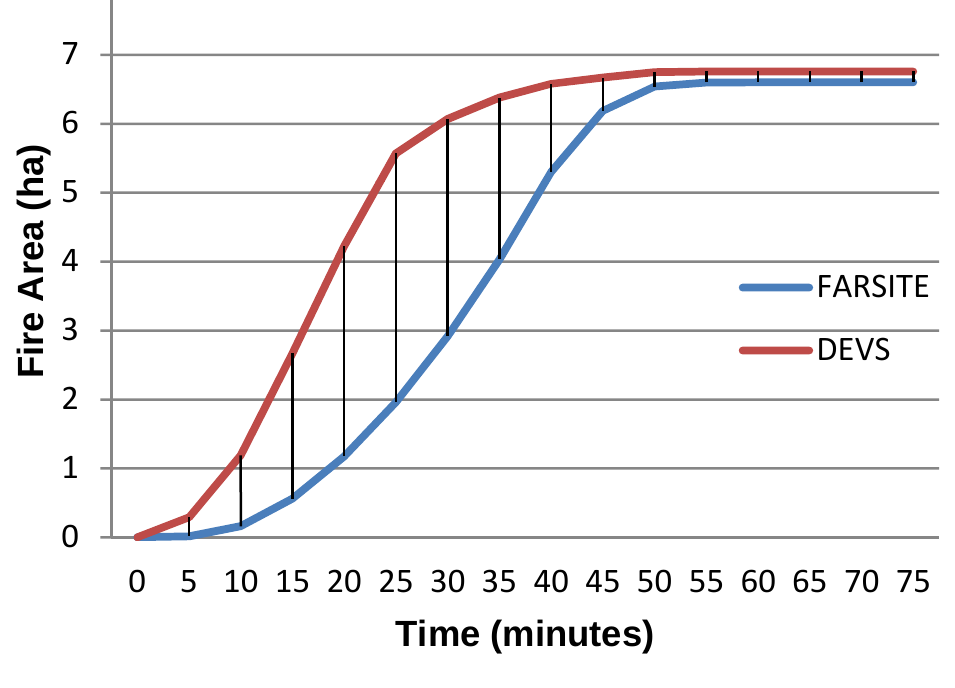}}
%.49\linewidth
\vspace{-0.3cm}
\caption{Fire area: FARSITE vs. DEVS.}
\label{fig:farsite-vs-devs}
}
\end{figure}

A comparison was done on the progression of fire area over time (see Figure~\ref{fig:farsite-vs-devs}). The result shows the likeness between both methods. Of course, the goal is not to mimic FARSITE software behavior, it is possible to fine-tune or parameterize our model so that it mimics FARSITE and obtain the same graph. However, the real purpose is to show the vitality of our approach. As you can see, the same scenario can{'}t be simulated with the forest cell space coupled model and we will get a message as {\tt Out of memory error}. A solution suggested is, either we divide this forest space into small areas or decrease the cell resolution (increase cell dimension) which may result in incorrect results and degrade accuracy.

As cited previously, this is a partial and preliminary validation. More validation is definitely required because of the many potential sources of error which can confuse the comparisons.  

These tests have underlined the necessity to consider errors associated with:
\begin{itemize}
\item  The spatial and temporal resolution of the inputs,
\item  The model input data (fuels and environmental parameters) and
\item  The nature of fire growth projections used for comparison in both methods (Huygens principle of wave propagation in FARSITE versus Rothermel spread equation in Behave package).
\end{itemize}

Logically, a wildfire spread simulation should be most accurate when using accurate data at high spatial and temporal resolutions. 

The acceptability of this statement is dependent on the spatial resolution required by the user and the resolution specified for the simulation. Thus, the sensitivity of fire simulation to the resolution, fire spread patterns generated (Huygens principle) and model of different fire input parameters remain to be verified. 

However, it should be noted that even FARSITE needs to be validated \cite{Finney2004}, the most important result of FARSITE tests to date has been proven that spread rates for all fuel models tended to be over predicted by the Rothermel spread equation. Different adjustments have been added in order to correct the observed error with real forest fire, but there is no way to guide settings for these adjustments. Consequently, future work needs to focus more on validating our approach using observed fire growth patterns in several conditions of weather and vegetation.

\section{Conclusion and Future Works}
\label{sec:conclusion}
This work attempts to foster the integration between two concepts: closure under coupling in cellular-DEVS and the forest fire spread to gain memory resource during simulation run. 

The study of the simulation of forest fire has increased considerably, it requires building simulation models that allows for system evolution in both time and space. 

The most common approaches to simulate fire spread are based either on vector approach, or on the cellular models. Cellular models simulate fire spread as a discrete process of ignitions across a regular division of the space in cells. However, dealing with high cell resolutions naturally defies efficient computer simulation. 

In this work, we tried to gain some performance in simulation run. This was achieved by converting the cell space into atomic model in order to eliminate inter-cell messages. This approach was initially deduced from the property closure under coupling of the DEVS formalism. Some relevant results are presented and a comparison between our approach and the simulation by FARISTE software is given.
 
However, this implementation needs an additional work in verification and validation (which we crossed an important step) to confirm more again the vitality of this approach. The functional architecture needs to take into account more parameters since the fire spread process might become complicated. A model for these parameters must be coupled to our architecture in order to get more realistic results. Also we need to validate our approach by using observed fire growth patterns in several conditions of weather and vegetation. An issue that we hope explore in the near future.

\bibliographystyle{model1-num-names}
\bibliography{arxivV1}

\end{document}